\documentclass[twocolumn]{aastex6}

\shorttitle{Orbital Motion of Young Binaries}
\shortauthors{Schaefer et al.}

\bibpunct[]{(}{)}{;}{a}{}{,}

\begin{document}

\title{Orbital Motion of Young Binaries in Ophiuchus and Upper Centaurus-Lupus}

\author{G. H. Schaefer\altaffilmark{1}, L. Prato\altaffilmark{2}, \& M. Simon\altaffilmark{3}}

\altaffiltext{1}{The CHARA Array of Georgia State University, Mount Wilson Observatory, Mount Wilson, CA 91023, USA (schaefer@chara-array.org)}
\altaffiltext{2}{Lowell Observatory, 1400 West Mars Hill Road, Flagstaff, AZ 86001, USA}
\altaffiltext{3}{Department of Physics and Astronomy, Stony Brook University, Stony Brook, NY 11794, USA}

\begin{abstract}

We present measurements of the orbital positions and flux ratios of 17 binary and triple systems in the Ophiuchus star forming region and the Upper Centaurus-Lupus cluster based on adaptive optics imaging at the Keck Observatory.  We report the detection of visual companions in MML 50 and MML 53 for the first time, as well as the possible detection of a third component in WSB 21.  For six systems in our sample, our measurements provide a second orbital position following their initial discoveries over a decade ago.  For eight systems with sufficient orbital coverage, we analyze the range of orbital solutions that fit the data.  Ultimately, these observations will help provide the groundwork toward measuring precise masses for these pre-main sequence stars and understanding the distribution of orbital parameters in young multiple systems.

\end{abstract}


\section{Introduction}

The orbits of binary stars provide a way to measure the dynamical masses of the stellar components.  Recent work on measuring the masses of pre-main sequence (PMS) stars has led to significant improvements in the evolutionary models computed at young stellar ages \citep[e.g.,][]{torres13,paxton11,baraffe15,feiden16,simon17}.  In this paper, we focus on measuring the visual orbits of 15 binary stars in the Ophiuchus star forming region that were initially resolved as binaries 10$-$20 yr ago \citep[e.g.,][]{ghez93,reipurth93,simon95,ageorges97,aspin97,koresko02,barsony03,ratzka05,correia06,mccabe06}.  Several of these systems have not had measurements of their positions published since the time of their discovery.  We also include two new detections of binaries in the Upper Centaurus-Lupus (UCL) cluster; these stars were initially included in a survey to detect tertiary companions in a sample of PMS spectroscopic binaries (Prato et al.\, in prep).  
If the distance is known, a visual binary orbit provides a measurement of the total system mass.  At the distance of the Ophiuchus star forming region \citep[$\sim$ 130 pc;][; and references therein]{cheetham15,mamajek08}, binaries resolvable through adaptive optics (AO) imaging typically have periods on the order of decades \citep[e.g.,][]{schaefer06,schaefer14}.  Therefore, the AO measurements presented here provide the groundwork for measuring high precision orbital parameters and dynamical masses in the future.  In order to derive individual masses of the component stars, additional information is required, such as measurements of the astrometric motion of the components around their center of mass \citep{duchene06,schaefer12,kohler13,kohler16} or the radial velocity variations in a double-lined spectroscopic binary \citep{boden05,torres12,simon13,lebouquin14}.  

In addition to providing dynamical masses as anchor points for the evolutionary tracks, the distribution of orbital parameters (eccentricities, semi-major axes) and mass ratios in binary systems provide clues to understanding the formation of binary stars \citep[e.g.,][]{duchene13,reipurth14}.  Moreover, about seven of the systems studied in this paper are part of triple or higher order multiple systems.  The relative alignment of the orbits between the inner and outer pairs can probe the initial conditions of star formation \citep{fekel81,sterzik02,reipurth14,tokovinin14}.  A similar assessment can be done by comparing the binary orbital plane with the orientation of circumstellar or circumbinary disks in the system \citep[e.g.,][]{kellogg17}.  The frequency of coplanar or misaligned systems can provide insights into the formation of the multiple systems and their dynamical evolution over time.

Based on our AO observations and previous measurements available in the literature, we provide an overview on the status of the orbital motion and analysis for each binary and triple system that was observed.  We also indicate which systems show variability in their flux ratios.

\section{Adaptive Optics Observations}

\subsection{Target Selection}

In the Ophiuchus star forming region, we selected previously known binary stars with separations in the range of 40--350 mas and $R$-band magnitudes within the sensitivity limits of the Keck AO system.  The range of separations was selected so that the binaries could be resolved through AO imaging at Keck and would be expected to show a change in orbital motion since the time of their last observation.  The observations presented in this paper do not represent a complete sample of binaries that meet these selection criteria.

Table~\ref{tab.targets} lists the binary stars that we observed, along with their right ascension, declination, distance, region, and spectral type.  We assigned distances to each target based on their association with nearby star forming regions.  \citet{ortizleon17} measured the parallaxes for 12 stars in the Ophiuchus dark cloud Lynds 1688 and for three stars in the eastern streamer Lynds 1689.  For our targets that are positionally coincident with Lynds 1688 and 1689, we quote the average distances measured by Ortiz-Leon et al.\ of 137.3 $\pm$ 1.2 pc and 147.3 $\pm$ 3.4 pc, respectively.  One of our targets, ROXs 47A (DoAr 51) in Lynds 1689, has a direct parallax of 143.2 $\pm$ 1.0 pc measured by Ortiz-Leon and we quote that value for the system.  For Ophiuchus targets that lie in neither Lynds 1688 nor Lynds 1689, we use the distance of 130 pc adopted by \citet{cheetham15}.  For MML 50 and 53 we use the mean distance of 140 $\pm$ 2 pc to UCL determined by \citet{dezeeuw99}.  Revised distances for each system will be available in the near future from the GAIA mission \citep{gaia16}.

\subsection{Keck NIRC2 Observations}

\begin{figure*}
	\plotone{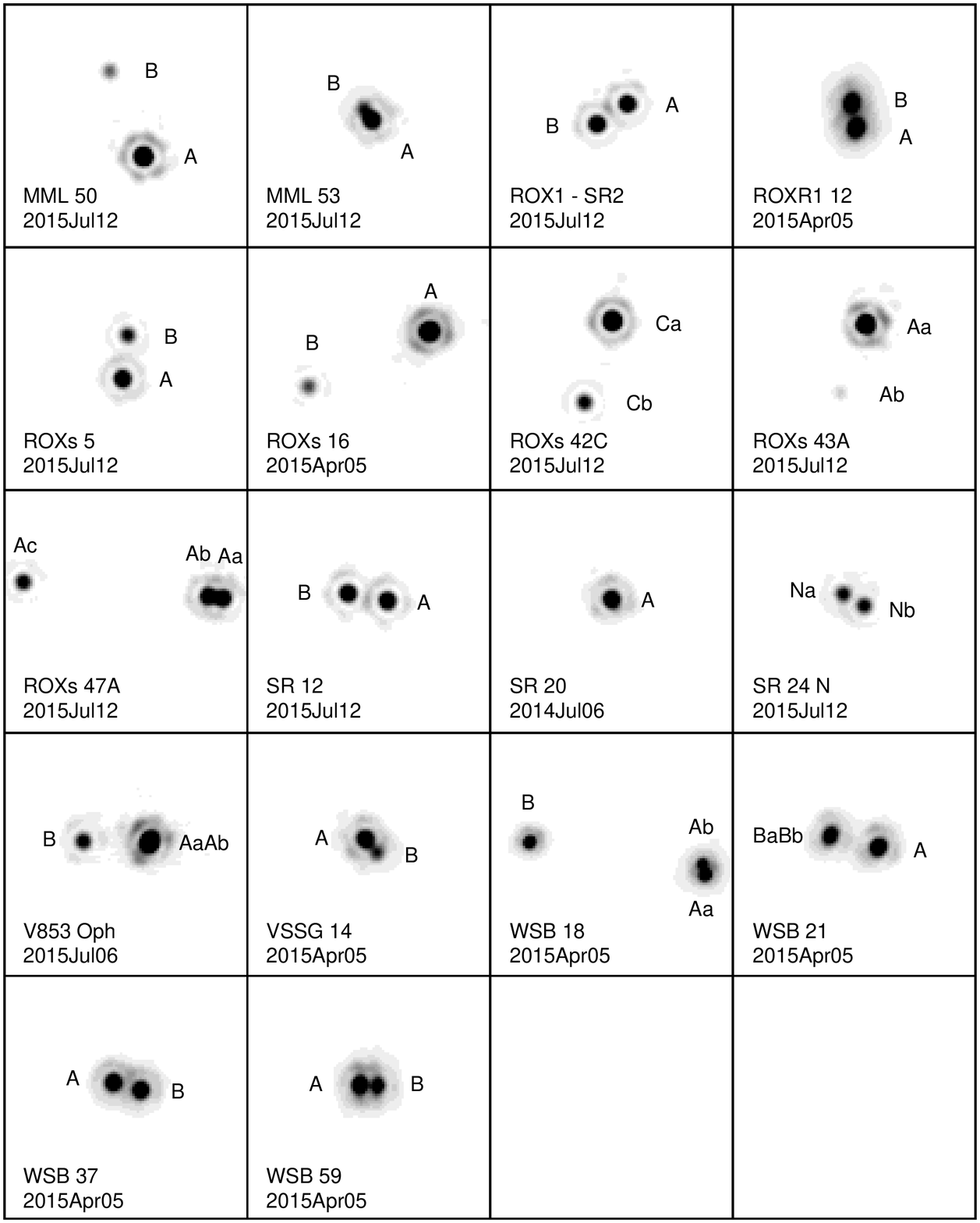}
	\caption{Keck NIRC2 AO images of PMS multiple systems in the Ophiuchus and UCL star forming regions.  Each panel is $\sim$1$''$ wide, except for the WSB 18 panel which is $\sim$1.5$''$ wide.  North is up and east is to the left.  Wide companions with separations larger than 1$''$ are not shown.  For systems where the components are nearly equal brightness, the identification of the A and B components might vary in the literature.  In most panels, the flux scale saturates at 0.4 times the maximum value, except for MML 50, ROXs 16, ROXs 42C, ROXs 43A, and V853 Oph which saturate at 0.2 times the maximum to bring out the faint companions.}
\label{fig.images}
\end{figure*}

We observed the sample of binary stars using the near-infrared camera NIRC2 \citep{wizinowich00} on the 10-m Keck II Telescope at the W.\ M.\ Keck Observatory.  We used the natural guide star AO system on all nights except for UT 2015 April 5 when we used the laser guide star.  MML 50 and MML 53 were observed using natural guide star AO on all nights (including UT 2015 April 5).   All images were taken with the narrow-field camera in NIRC2 which has a field of view of 10$\arcsec$.   Table~\ref{tab.aolog} provides an observing log which lists the UT date and time of observation, target name, filter, AO rate, integration time per exposure, and number of images taken in each filter.  Each image consisted of 10 coadded exposures.  We obtained sets of 5$-$12 images dithered across the detector.   We flatfielded the images using dark-subtracted dome flats and removed the sky background by subtracting pairs of dithered images.  Figure~\ref{fig.images} shows example images of the multiples in our sample with the components identified.  

%0.4: MML 53, ROX 1, ROXR1 12, ROXs 5, ROXs 47A, SR 12, SR 20, SR24 N, VSSG 14, WSB 18, WSB 21, WSB 37, WSB 59 [13]
%0.2: MML 50, ROXs 16, ROXs 42C, ROXs 43A, V853 Oph [5]

We computed the separation, position angle, and flux ratio of the components in the multiple systems using point-spread function (PSF) fitting following the same methods described in \citet{schaefer14}.  For triple systems, we used the wide component as a simultaneous PSF to model the close pair.  For wide binaries where the Airy rings of the PSF for each component did not overlap ($\gtrsim 200$ mas), we used the primary component as the PSF to measure the relative separation of the secondary.  Ordinarily, for close binaries with overlapping PSFs and no wide component in the field of view, we typically use a separate observation of a single star PSF reference observed either immediately before or after the target using the same AO frame rate.  However, because the targets are at southern declinations and were observed at moderate airmasses ($z$ = 1.4$-$2.0), the AO correction between nearby targets was not always stable, leading to mis-matched PSF shapes.  The variability in the shape of the central core and structure in the Airy ring of the PSF can be seen by comparing images of different binaries observed on the same night in Figure~\ref{fig.images}.  To overcome this difficulty, we created an ``effective PSF'' (ePSF) directly from the blended binary components.  

\begin{figure}
	\plotone{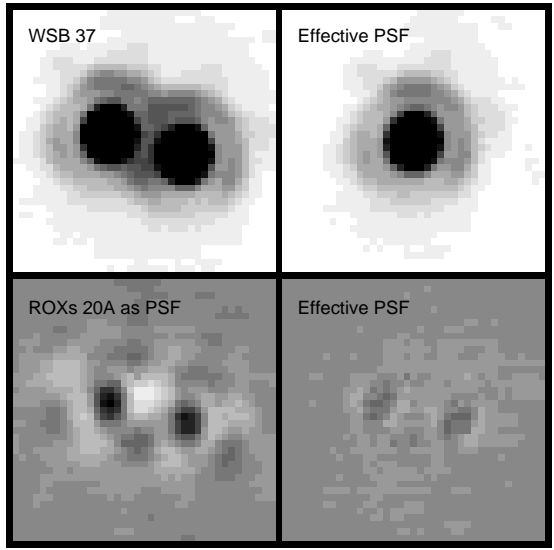}
	\caption{Top left: Image of the binary WSB 37 on UT 2015 Apr 5.  The binary is separated by 117 mas.  Top right: Effective PSF created by aligning and coadding the PSFs from the two components after rejecting pixels with the highest number of counts.  Bottom: Residuals between WSB 37 and the model binary created using a separate image of ROXs 20A as the PSF (left) and using the effective PSF (right).  The two residual panels are plotted on the same scale, showing the improvement in the binary fit when using the effective PSF.}
\label{fig.epsf}
\end{figure}

To create the ePSF, we cross-correlated the central cores of the PSFs from each component to align the images, scaled the fluxes to match their brightnesses, and rejected pixels with the highest counts from the averaged PSF.  This process eliminates the blended side of the PSF from each component, providing the PSF of an effectively ``single'' star as shown in Figure~\ref{fig.epsf}.  In each case where we used an ePSF (indicated in the last column of Table~\ref{tab.aolog}), we compared the positions with those computed from using a separate observation of a single PSF star.  In all cases, the $\chi^2$ between the data and binary model were lower when using the ePSF.  For binaries with separations $\lesssim$ 70 mas (i.e., MML 53 and VSSG 14), the cores of the components overlapped too much, so we could not create a suitable ePSF and had to use a separate observation of a single star as the PSF.  We tested the reliability of creating an ePSF using the triple star EM* SR 24.  For this system, we used SR 24S as a simultaneous PSF to model the $\sim$ 100 mas close pair SR 24 Na,Nb.  We also created an ePSF directly from the components in the close pair.  The positions derived for the close pair were consistent within 0.90 mas on UT 2014 July 06 and 0.39 mas on UT 2015 July 22, both within the 1\,$\sigma$ uncertainties of the two methods.

\begin{figure*}
\includegraphics[height=3.5cm]{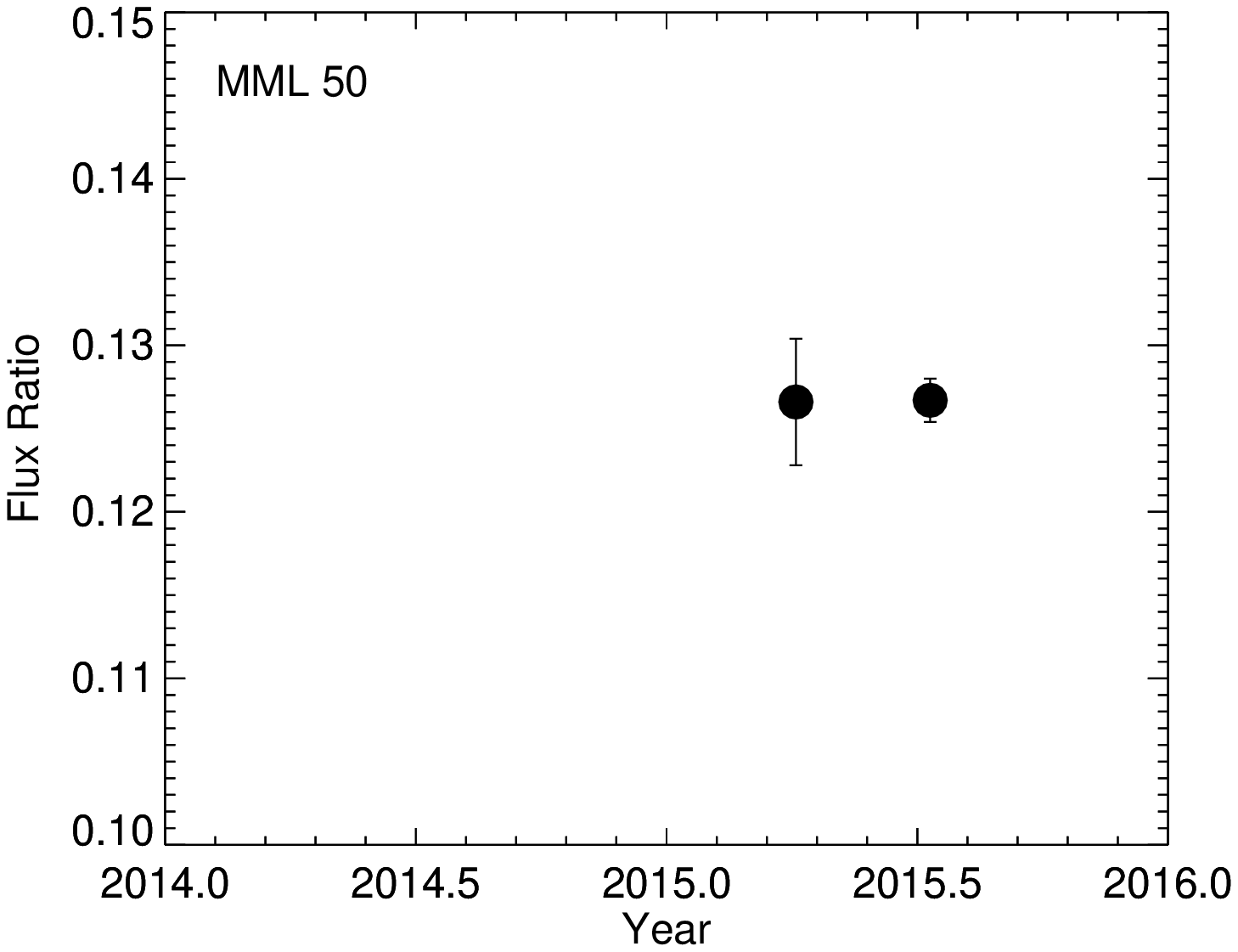}
\includegraphics[height=3.5cm]{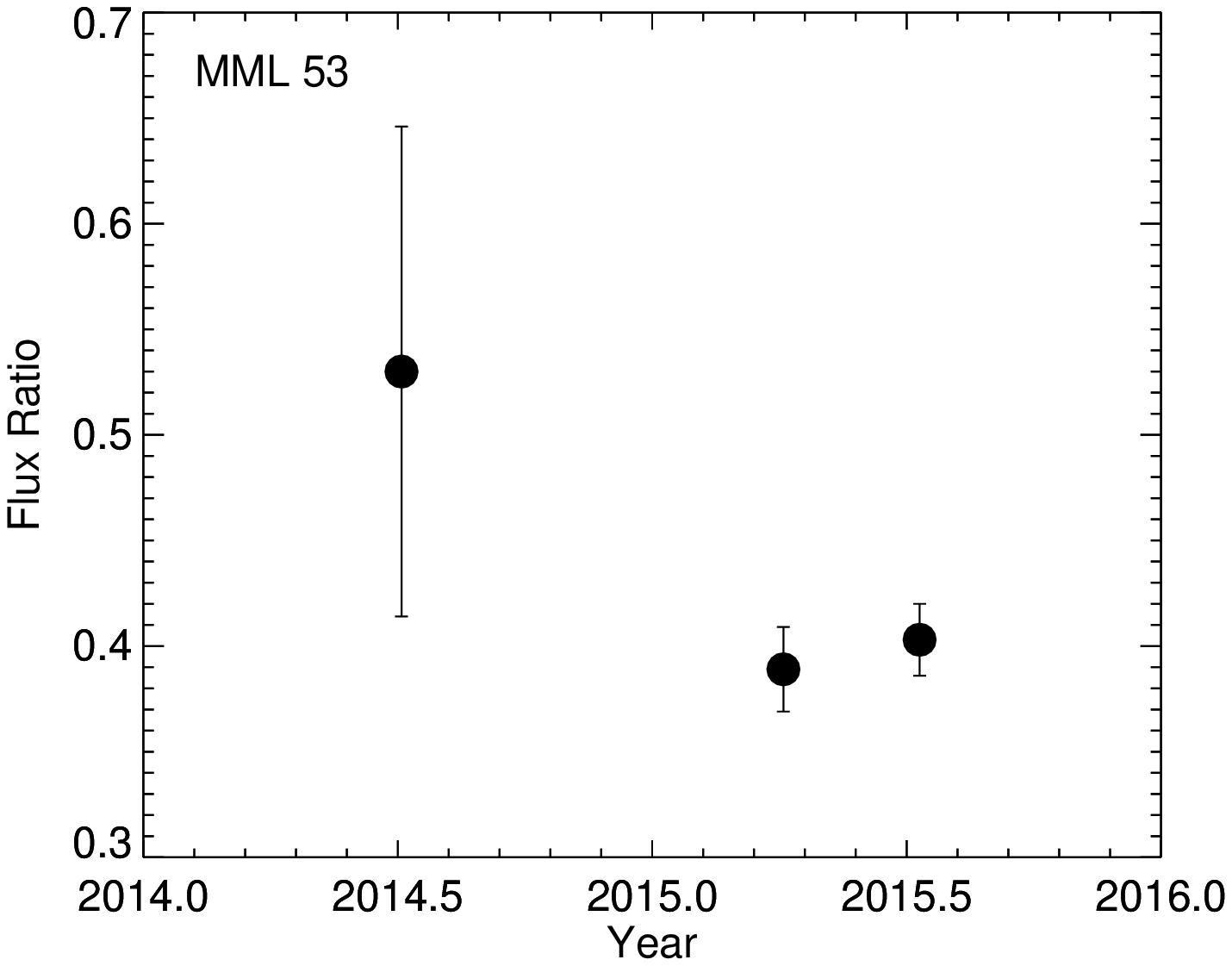} 
\includegraphics[height=3.5cm]{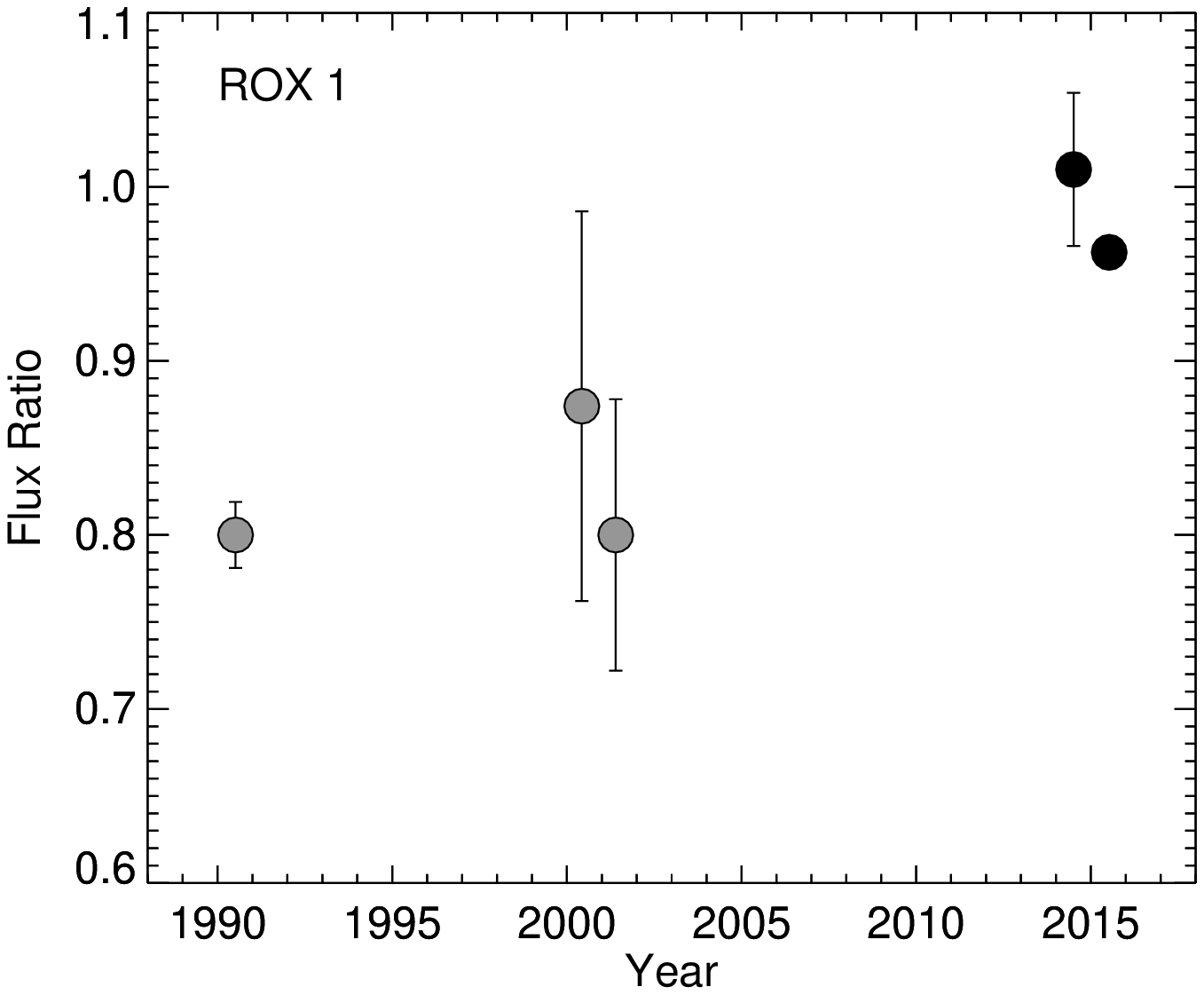} 
\includegraphics[height=3.5cm]{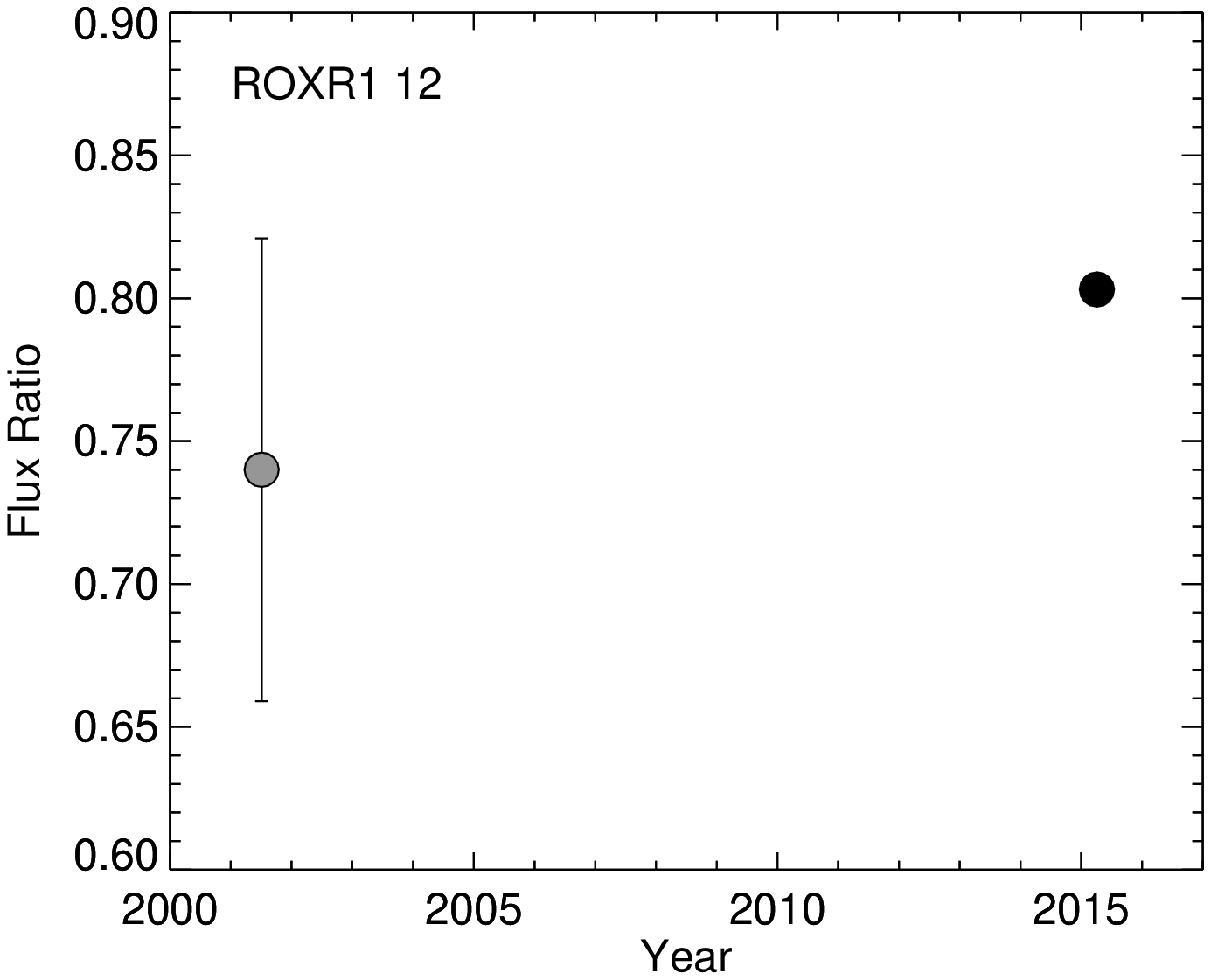} \\  
\includegraphics[height=3.5cm]{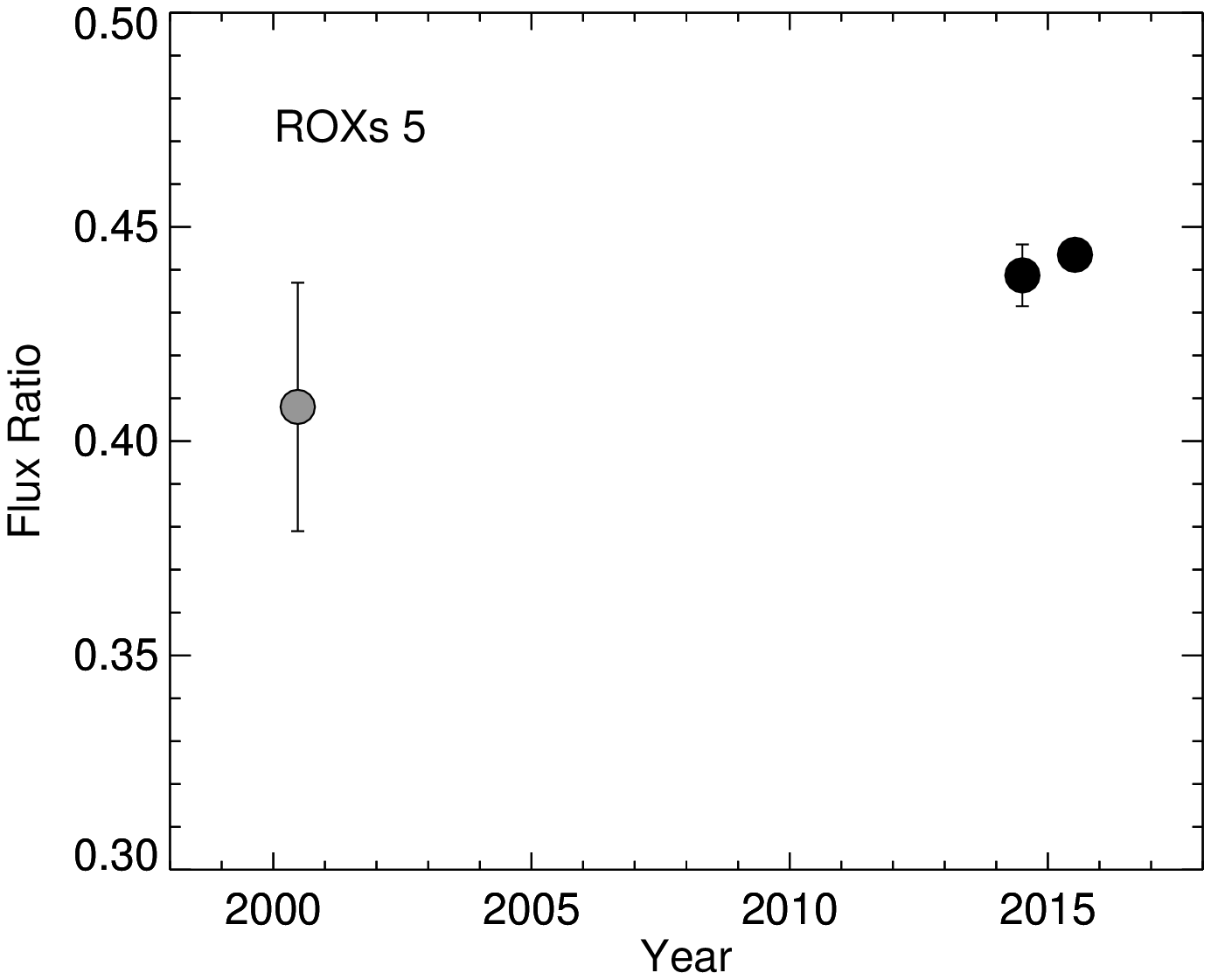}
\includegraphics[height=3.5cm]{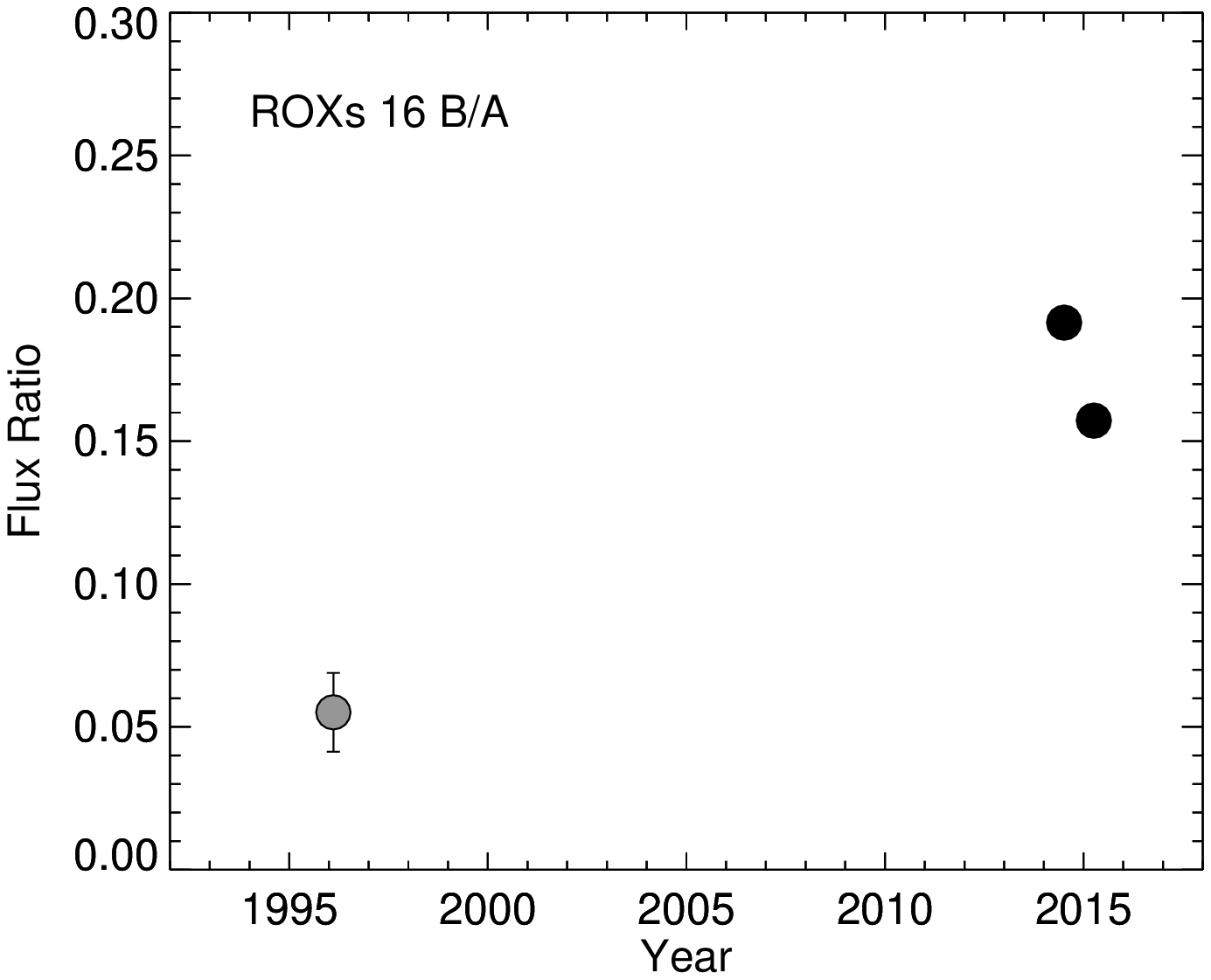}
\includegraphics[height=3.5cm]{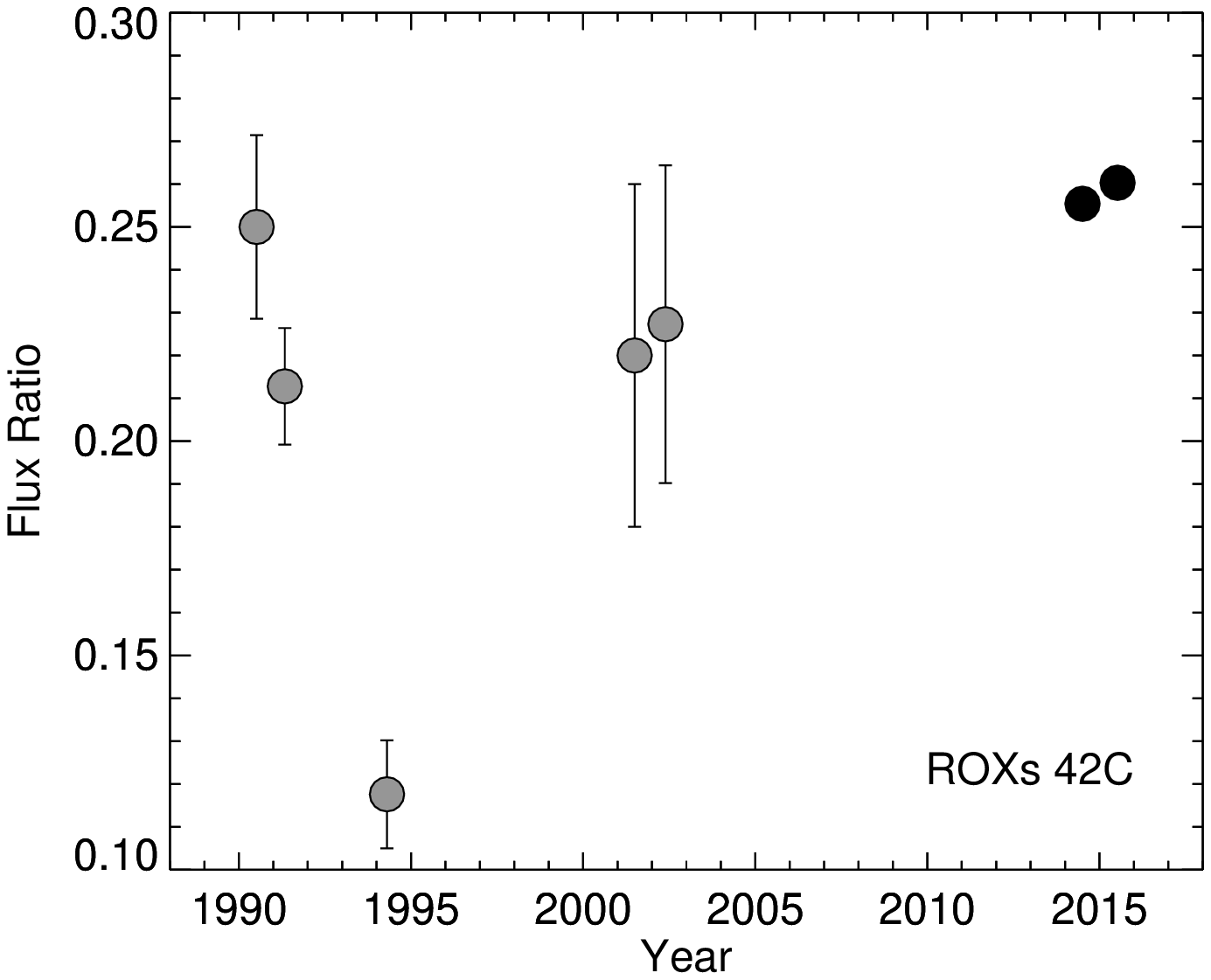} 
\includegraphics[height=3.5cm]{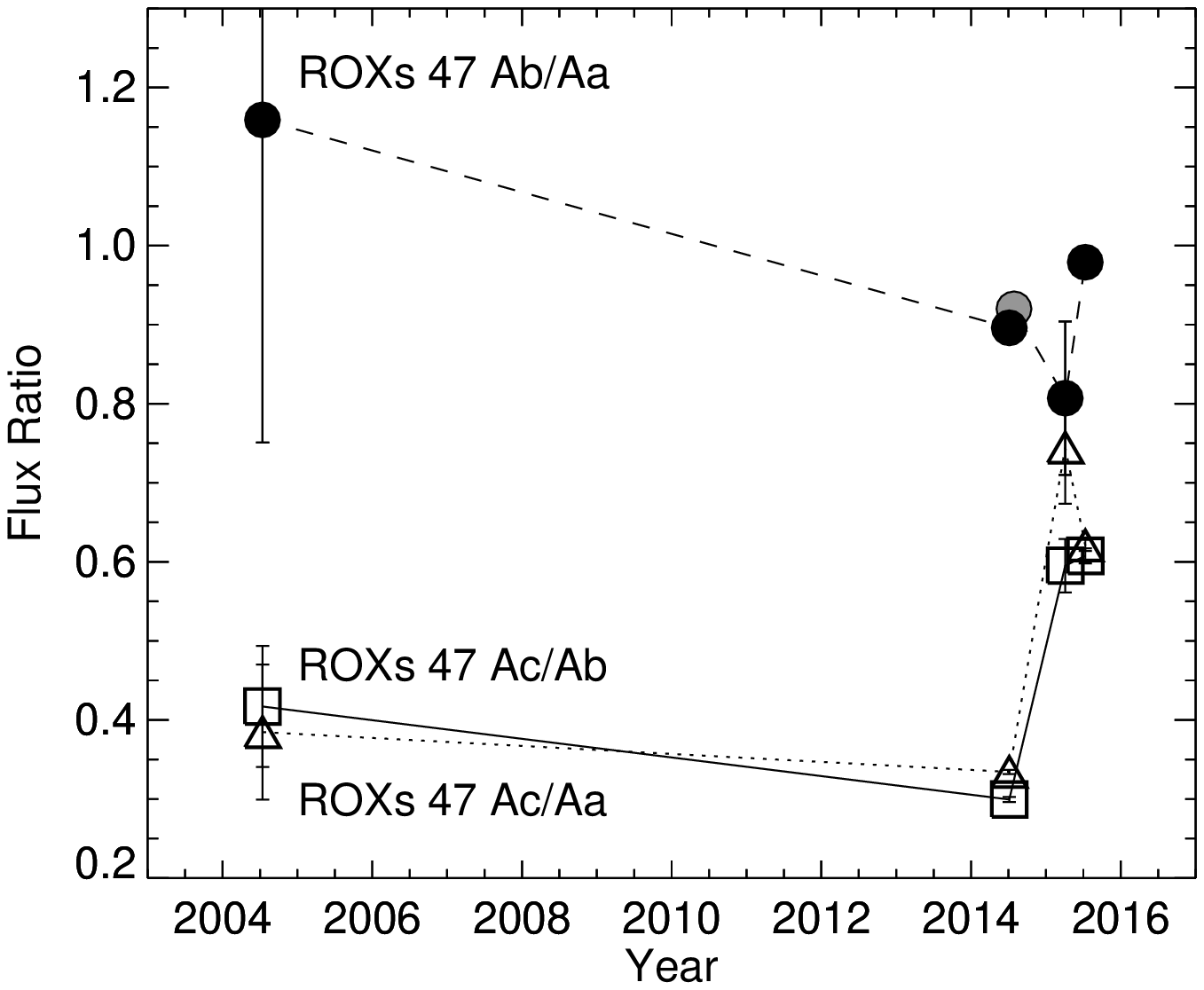} \\ 
\includegraphics[height=3.5cm]{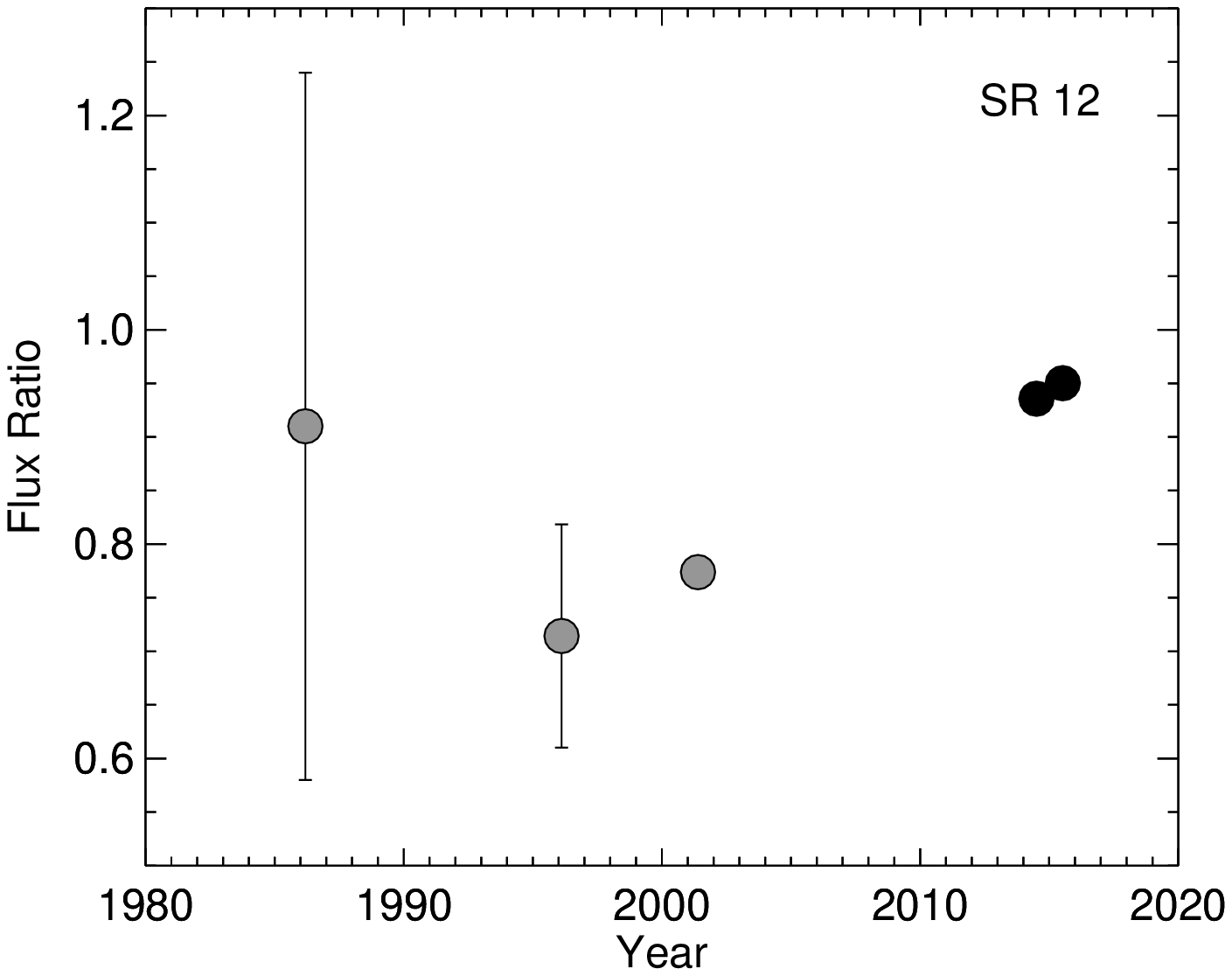} 
\includegraphics[height=3.5cm]{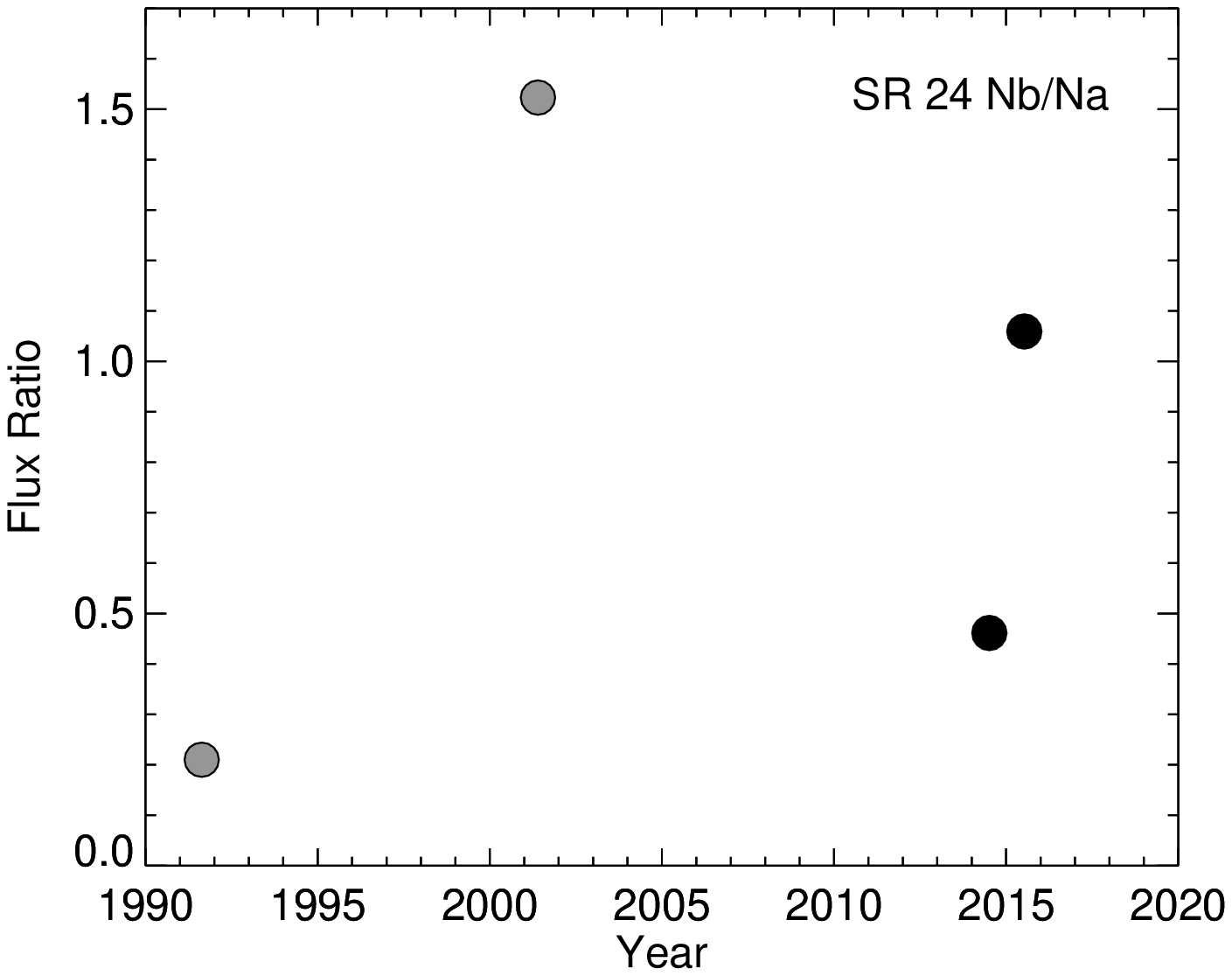} 
\includegraphics[height=3.5cm]{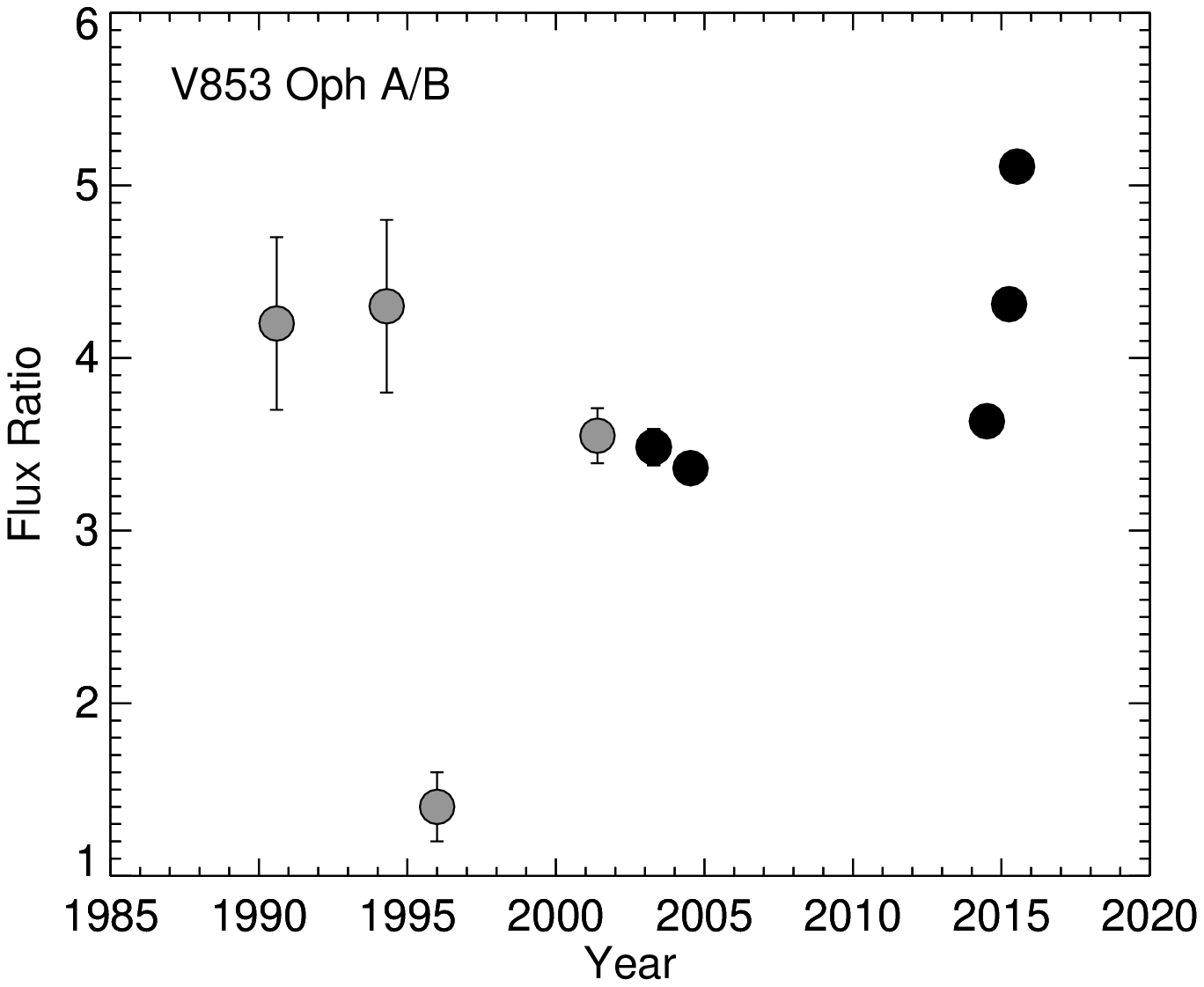}
\includegraphics[height=3.5cm]{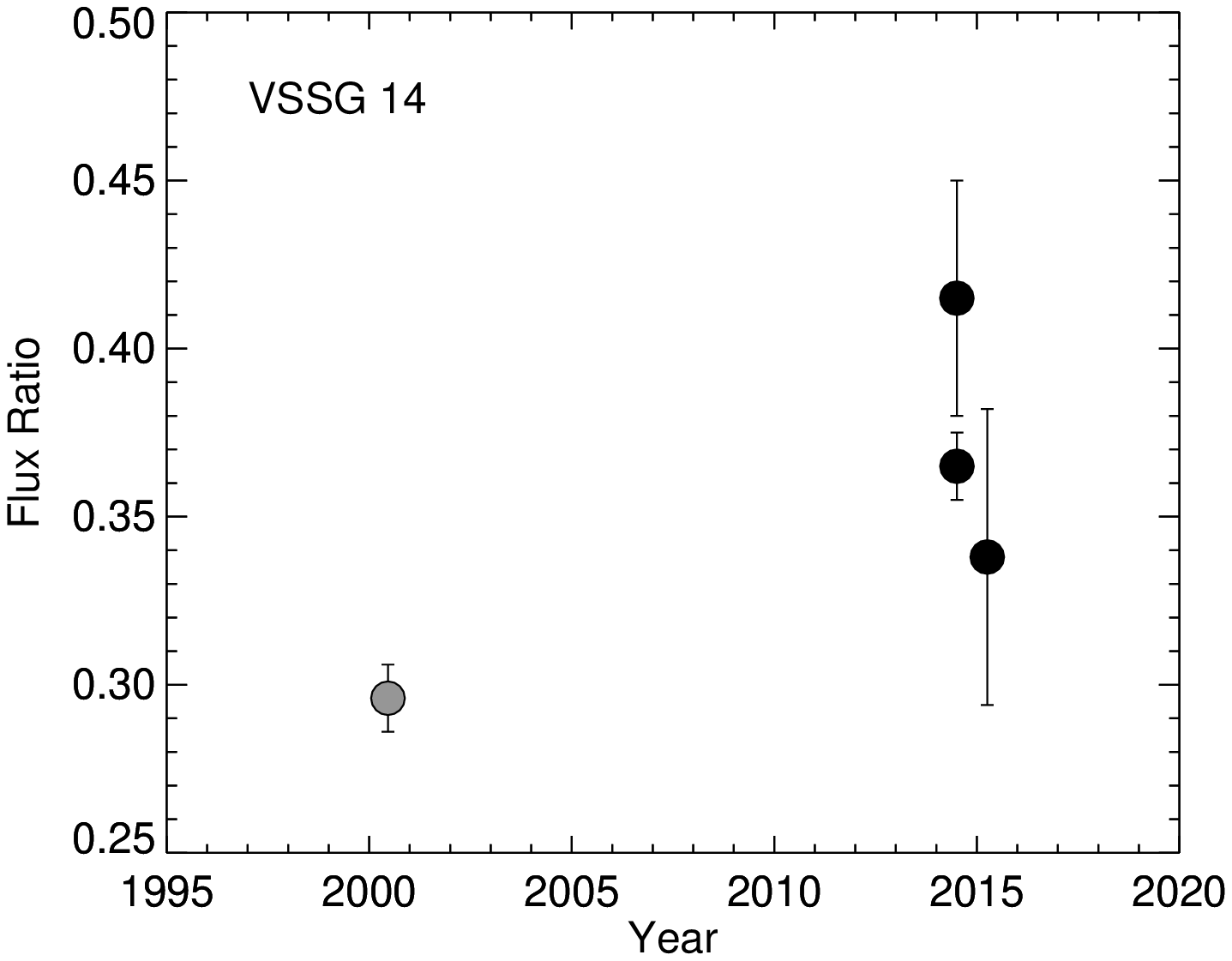} \\
\includegraphics[height=3.5cm]{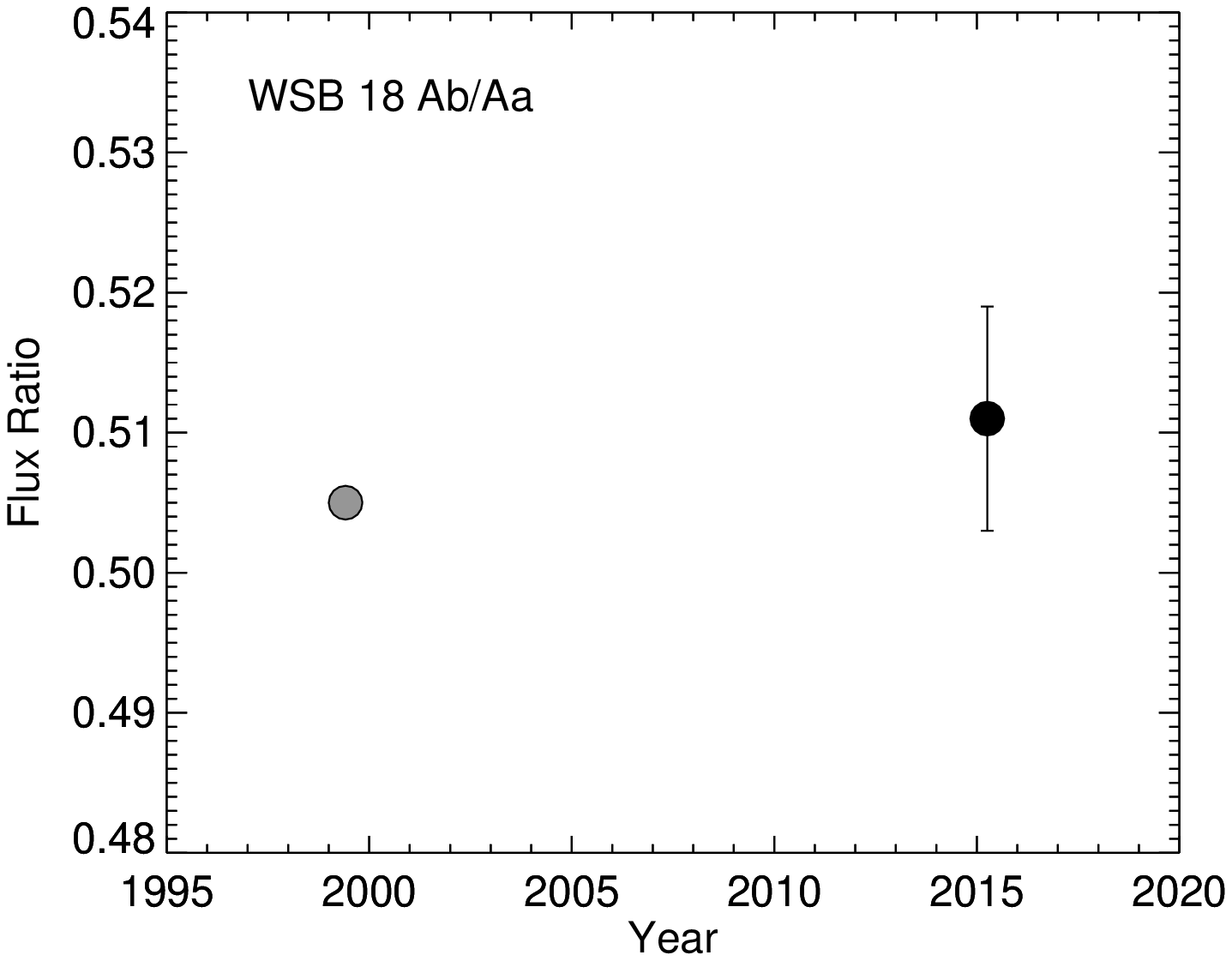}
\includegraphics[height=3.5cm]{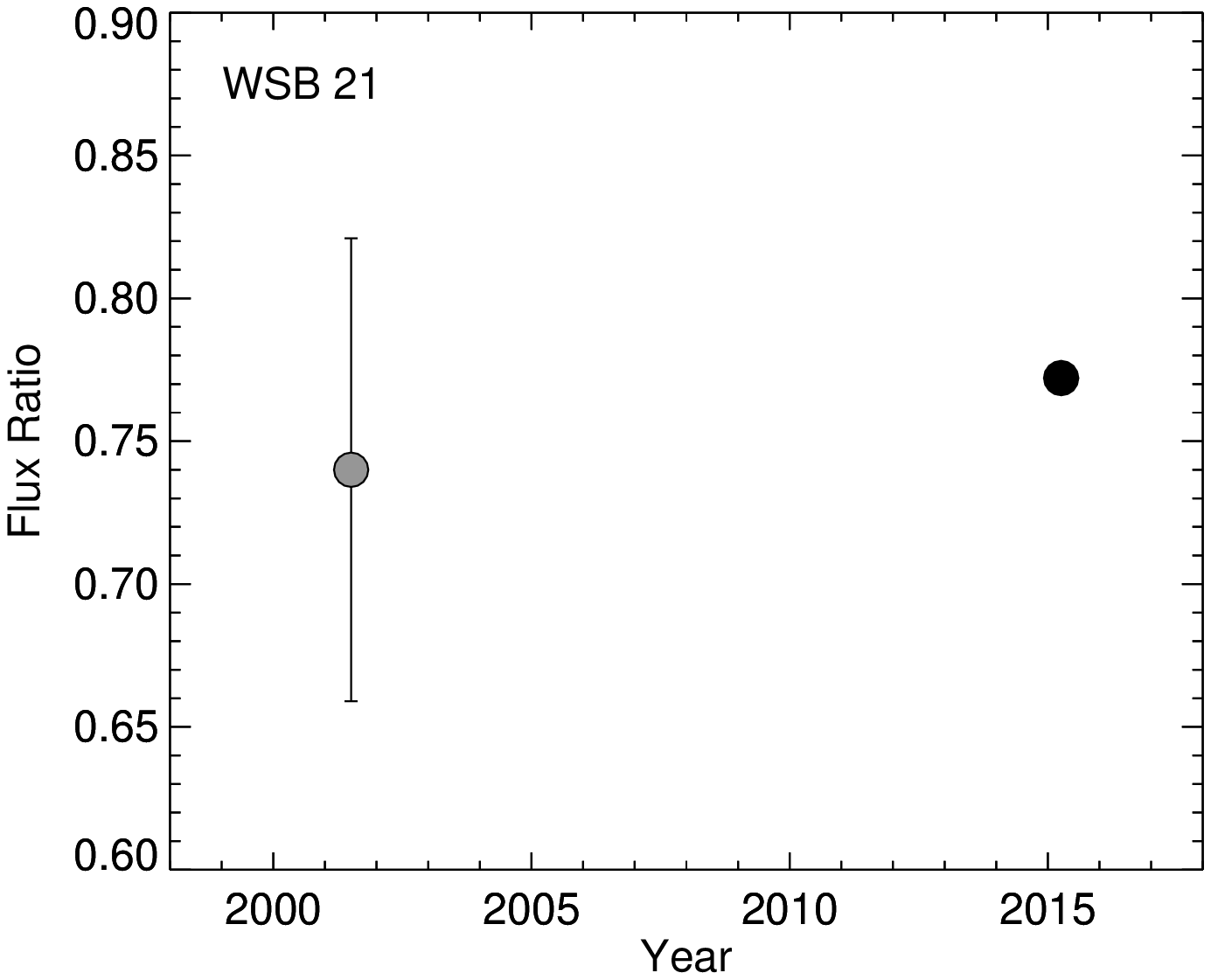}  
\includegraphics[height=3.5cm]{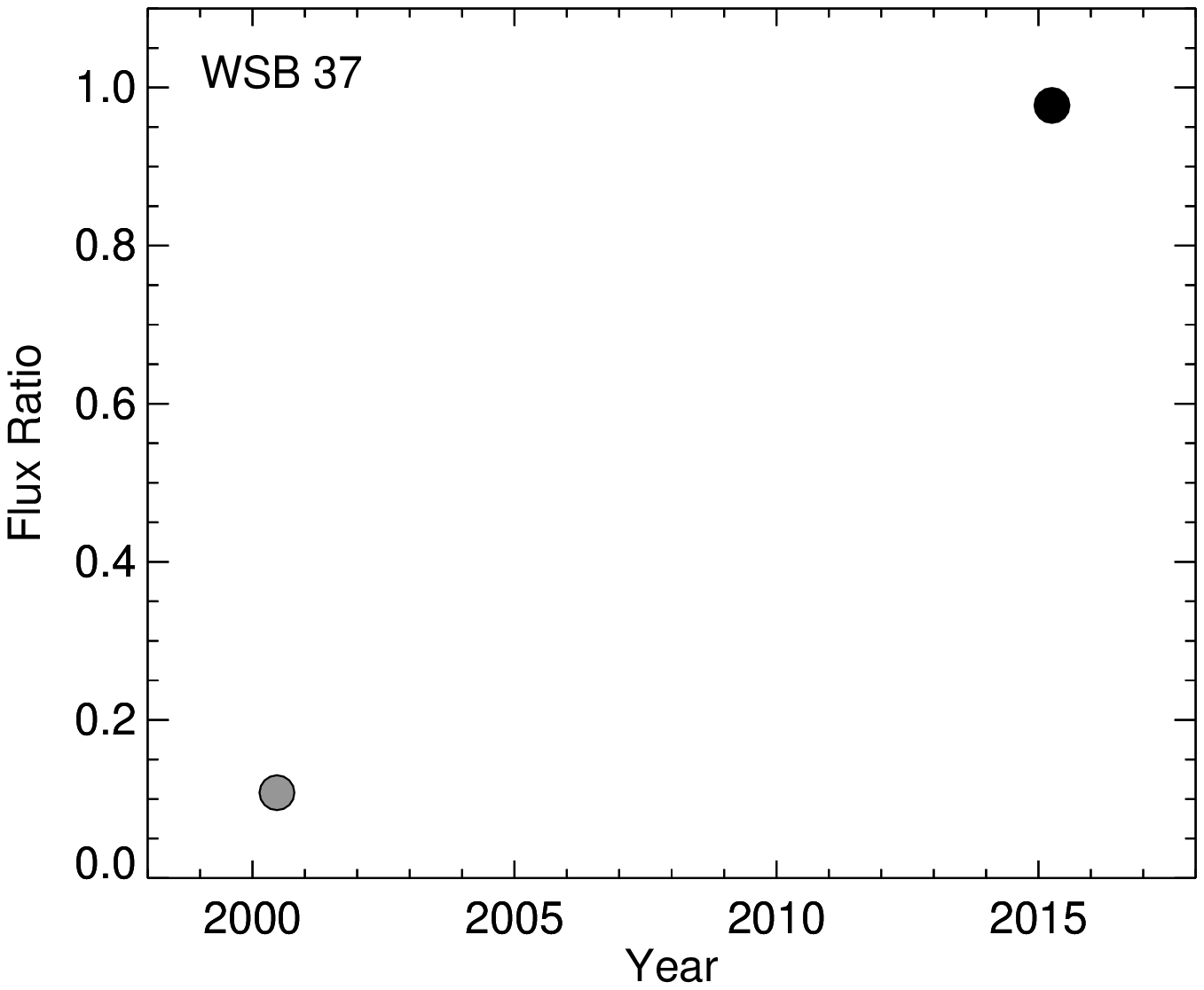} 
\includegraphics[height=3.5cm]{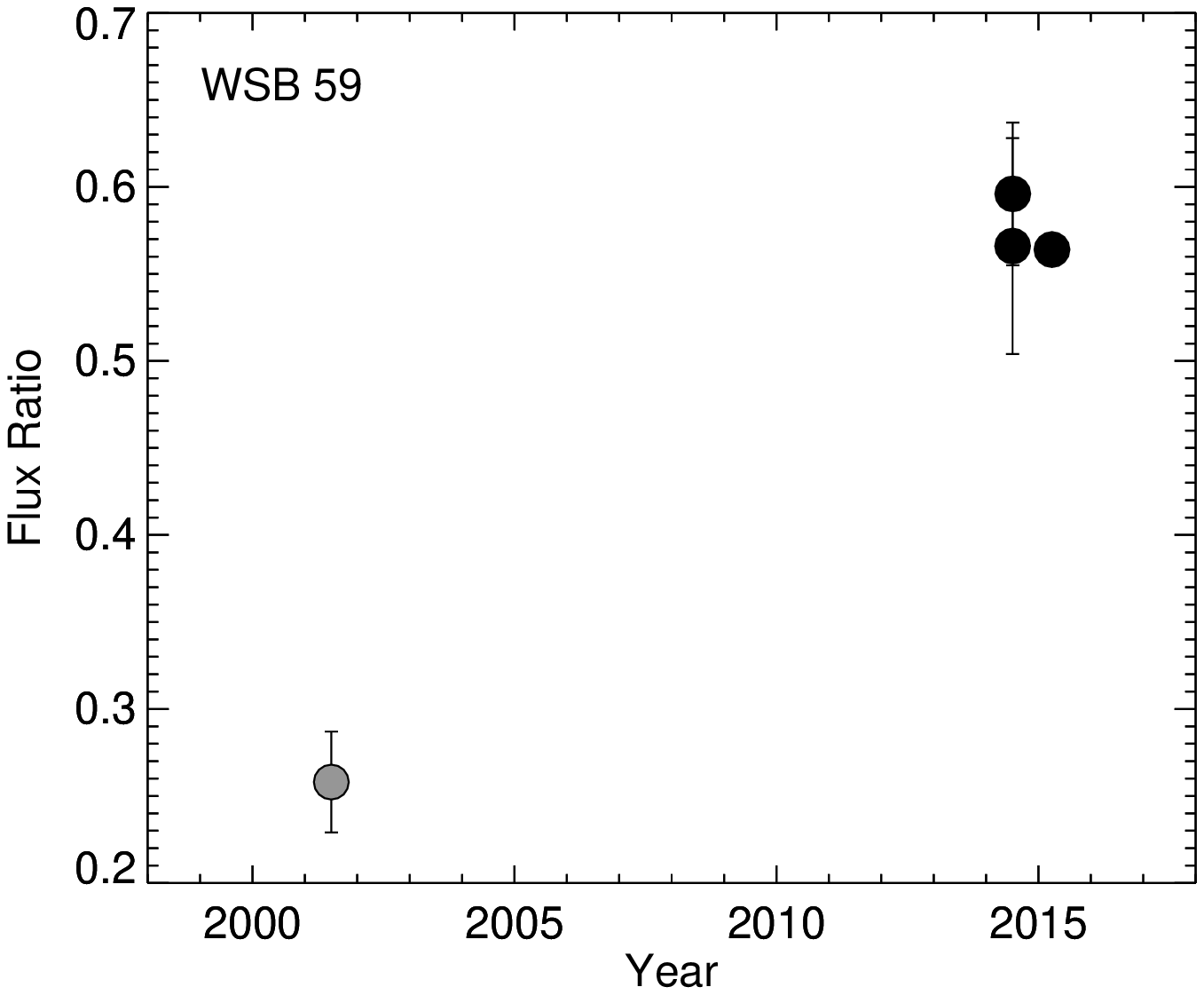}
	\caption{$K$-band flux ratios for each system as a function of time.  The black circles show our AO measurements while the gray circles show previously published measurements from the literature (see references in notes on individual systems in Section~\ref{sect.orbit}).}
\label{fig.fratio}
\end{figure*}

We corrected the binary positions using the geometric distortion solutions published by \citet{yelda10} prior to the optical realignment of the AO system on 2015 April 13 and by \citet{service16} after the realignment.  For the earlier AO observations, we used a plate scale of 9.952 $\pm$ 0.001 mas~pixel$^{-1}$ and subtracted $0\fdg252 \pm 0\fdg009$ from the raw position angles to correct for the orientation of the camera relative to true north.  After 2015 April 13, we used a plate scale of 9.971 $\pm$ 0.004 mas~pixel$^{-1}$ and subtracted $0\fdg262 \pm 0\fdg020$ from the measured position angles.  Table~\ref{tab.sepPA} presents the Julian year, binary separation, position angle measured east of north, and flux ratios measured in each filter.  The positions were averaged over the measurements from individual frames in the $H$ and $K$ bands, and uncertainties were computed from the standard deviation.  These filters provide a good compromise between angular resolution and PSF stability.  For systems with more than two components, we list the measurements for each pair of stars separately.  In Figure~\ref{fig.fratio} we show how the $K$-band flux ratios change over time.

\section{Analysis of Orbital Motion}
\label{sect.orbit}

We combined our AO measurements of the binary positions with previously published values from the literature.  In several cases we found no published reports of additional measurements of the binary separations since the time of their discovery over 10$-$20 yr ago.  With such limited orbital coverage, it was not possible to compute orbital solutions.  For binaries with enough measurements to begin mapping the curvature of the orbit, we assessed the range of possible orbital solutions using the statistical approach outlined by \citet{schaefer06,schaefer14}.  In general, we searched for possible orbital solutions out to periods of 500 years, eccentricities from 0 to 0.99, and times of periastron passage that covered the full range of the orbital periods.  In three cases where the binaries have sufficient orbital coverage (ROXs 47A, EM* SR 20, and EM* SR 24), we computed preliminary values for the orbital period ($P$), time of periastron passage ($T$), eccentricity ($e$), angular semi-major axis ($a$), inclination ($i$), position angle of the line of nodes ($\Omega$), and the angle between the node and periastron ($\omega$).  We discuss the results for each system in more detail below.

\subsection{MML 50}

MML 50 is a young star in the UCL cluster \citep{wichmann97,mamajek02}.  It was selected as a PSF star for MML 53, but was discovered to be binary with a separation of 376 mas with a flux ratio of 0.10 in the near infrared.  The system could be a triple because \cite{white07} identified the brighter component as a double-lined spectroscopic binary based on the presence of doubled lines.

\subsection{MML 53}

MML 53 is an X-ray active, lithium rich star in the UCL cluster \citep{wichmann97,mamajek02}.  It was identified as a double-lined spectroscopic binary by \citet{white07} based on the doubling of photospheric spectral lines observed in the visual.  The system was later discovered to be an eclipsing binary with a period of 2.1 days \citep{hebb10}.  Subsequently, a double-lined spectroscopic orbit was determined \citep{hebb11}.  A tertiary companion was detected on the basis of its spectral signature and a change in the systematic velocity of the eclipsing binary \citep{hebb10,hebb11}.

\begin{figure}
	\plotone{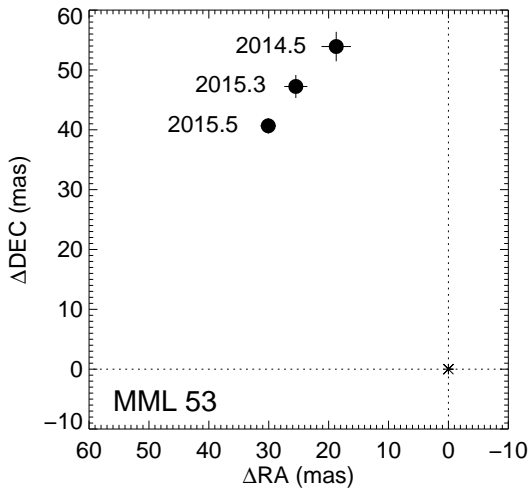}
	\caption{Orbital motion measured for MML 53 based on our Keck NIRC2 observations.}
\label{fig.mml53_orb}
\end{figure}

We resolved the tertiary companion at a separation of 57 mas using Keck NIRC2 AO on UT 2014 July 5.  We resolved this companion again on two nights in 2015.  Based on the effective temperatures of the three components derived by \citet{hebb10} and the near-infrared flux ratio of $\sim$ 0.4 for the spatially resolved pair, we expect that the eclipsing binary is the brighter component in our NIRC2 observations.  The tertiary companion shows significant orbital motion over two years, as shown in Figure~\ref{fig.mml53_orb}.  However, we have not yet mapped enough of the orbit to place meaningful constraints on the orbital parameters.

\subsection{ROX 1}

\begin{figure}
	\plotone{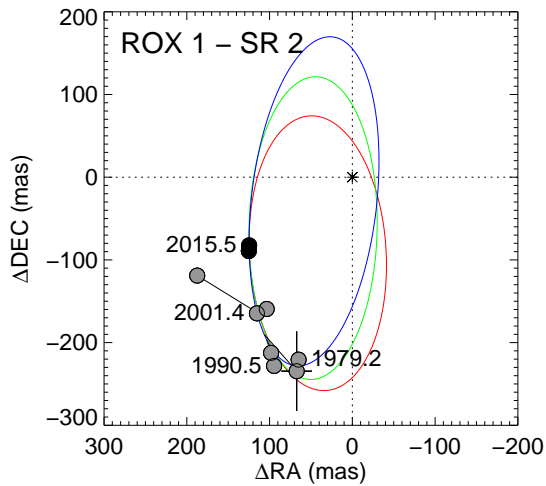}
	\caption{Orbital motion measured for ROX 1 (EM* SR 2) based on our AO observations (black circles) and published values from the literature \citep[gray circles;][]{heintz80,ghez93,ghez95,aspin97,ratzka05,mccabe06}.  Overplotted are three sample orbits with periods ranging from 110--130 yr and total masses of $\sim 1.9 M_\odot$ for a distance of 130 pc.}
\label{fig.rox1_orb}
\end{figure}

ROX 1 (EM* SR 2) was first detected as a binary through micrometer observations by \citet{heintz80}.  The binary was subsequently resolved through speckle imaging \citep{ghez93,ghez95,ratzka05,mccabe06} and shift-and-add techniques \citep{aspin97}.  We resolved the system twice in 2014--2015 using AO imaging.  Figure~\ref{fig.rox1_orb} shows that there is a surprising amount of scatter in the measured orbital motion and that determining a well-defined orbital period is not yet possible.  Given that the components are nearly equal brightness, we flipped some of the position angles by 180$^\circ$ to find a realistic set of orbital positions.  It is possible that our identification of the two components at each epoch is not correct; additional measurements in the future to further map the orbital motion should help resolve these possible discrepancies.  

A statistical analysis of orbital solutions that fit the existing data indicates that the orbital period is greater than 60 yr, with the 1\,$\sigma$ confidence interval extending out to the search range of 500 yrs.  In Figure~\ref{fig.rox1_orb} we plot examples of three orbits that fit the data.  These orbits were selected to produce a total mass of $\sim 1.9 M_\odot$ at a distance of 130 pc.  The approximate total mass was estimated by assuming two equal mass components, each with a spectral type of G9 \citep{torres06} and using the effective temperature and mass estimates computed by \citet{kraus07}.  Without the constraint on the total mass, the orbit fits tend to prefer highly eccentric, high mass orbits.

\subsection{ROXR1 12}

\begin{figure}
	\plotone{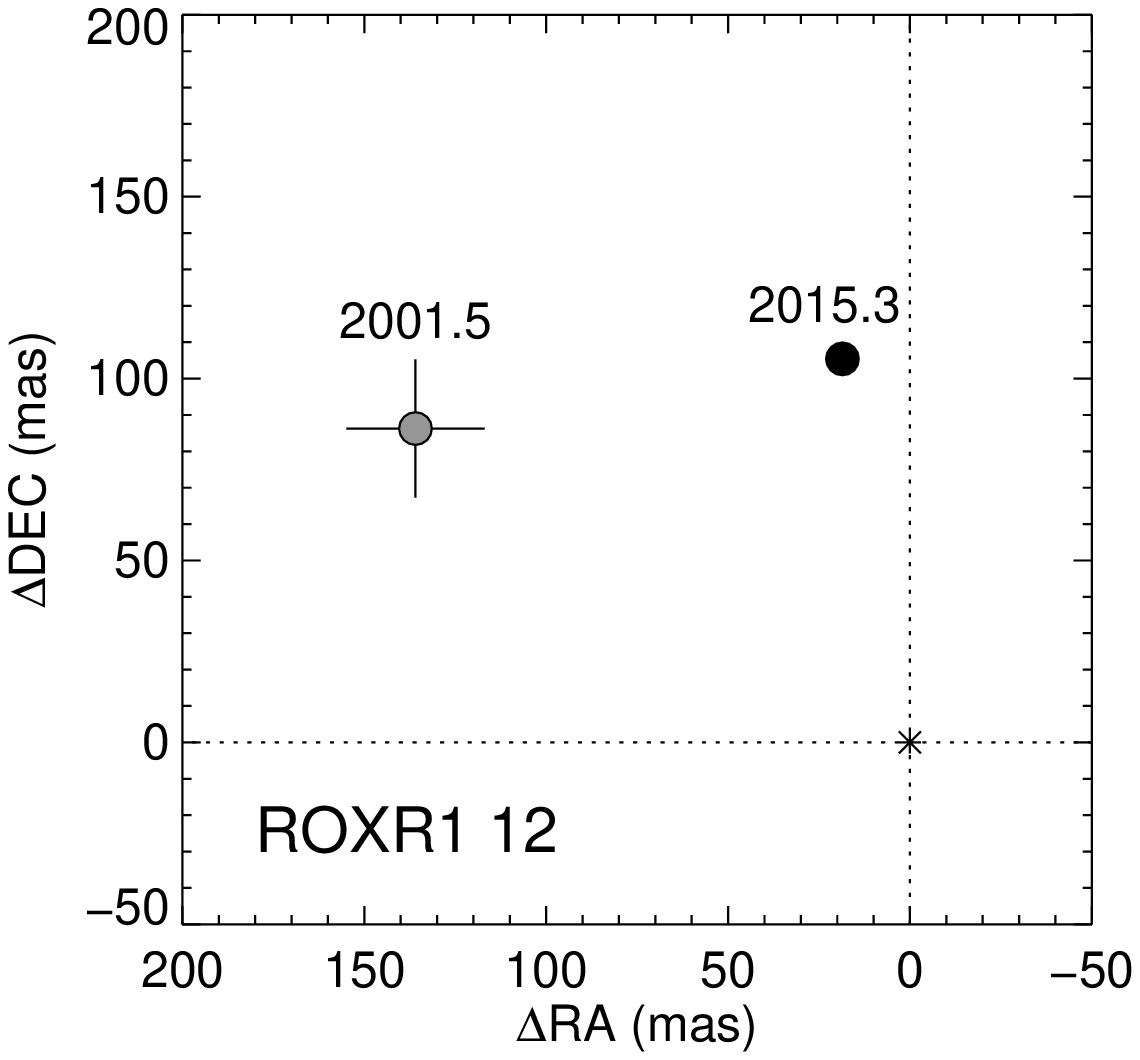}
	\caption{Orbital motion measured for ROXR1 12 based on our AO observations (black circle) and the speckle measurement from \citet[gray circle;][]{ratzka05}.}
\label{fig.roxr1_12_orb}
\end{figure}

ROXR1 12 was resolved as a binary through speckle observations by \citet{ratzka05} in 2001.  We resolved the system again using AO imaging in 2015.  Figure~\ref{fig.roxr1_12_orb} shows significant orbital motion between the two epochs, although with only two measurements, the coverage is not sufficient for fitting a binary orbit yet.

\subsection{ROXs 5}

\begin{figure}
	\plotone{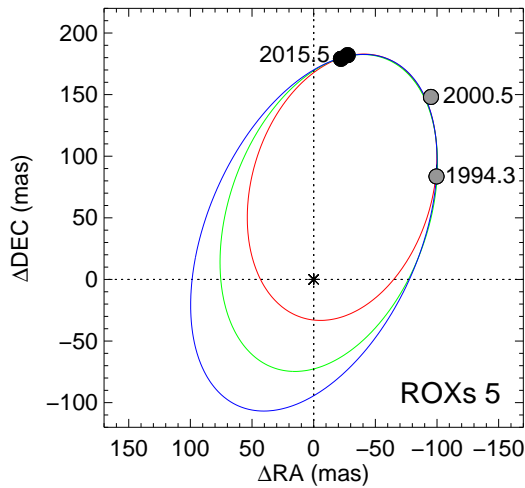}
	\caption{Orbital motion measured for ROXs 5 based on our AO observations (black circles) and published measurements from the literature \citep[gray circles;][]{ageorges97,ratzka05}.  We flipped the position angle of the speckle measurement by \citet{ageorges97} in 1994.3 by 180$^\circ$.  Overplotted are three sample orbits with periods ranging from 51--84 yr and total masses of $0.9 - 1.2 M_\odot$ for a distance of 130 pc.}
\label{fig.roxs5_orb}
\end{figure}

ROXs 5 was resolved as a binary through speckle observations by \citet{ageorges97} and \citet{ratzka05}.  We resolved the system two times using adaptive optics imaging at Keck.  Figure~\ref{fig.roxs5_orb} shows the orbital motion of ROXs 5.  The position angle of the speckle measurement by \citet{ageorges97} in 1994.3 has been flipped by 180$^\circ$.  At the published position angle of 130$^\circ$, a very eccentric family of orbits fit the data with a minimum total mass of 12.6 $M_\odot$ assuming a distance of 130 pc, inconsistent with the K7 spectral type of the system \citep{bouvier92}.  After flipping the position angle of the discovery measurement, a statistical analysis of orbital solutions that fit the data indicates that the orbital period is greater than 36 yr, with the 1\,$\sigma$ confidence interval extending out to the search range of 500 yr.  These orbital solutions provide a more realistic minimum mass of $\sim$ 0.87 $M_\odot$.  

In Figure~\ref{fig.roxs5_orb} we show examples of three orbits that fit the data.  When selecting the orbits to plot, we placed an upper limit of $\sim 1.2~M_\odot$ on the total system mass, determined by assuming two equal mass components, each with a spectral type of K7 \citep{bouvier92}, and using the table of effective temperatures and masses computed by \citet{kraus07}.  This removed highly eccentric, high mass orbits from the range of possible solutions.

\subsection{ROXs 16}

\begin{figure}
	\plotone{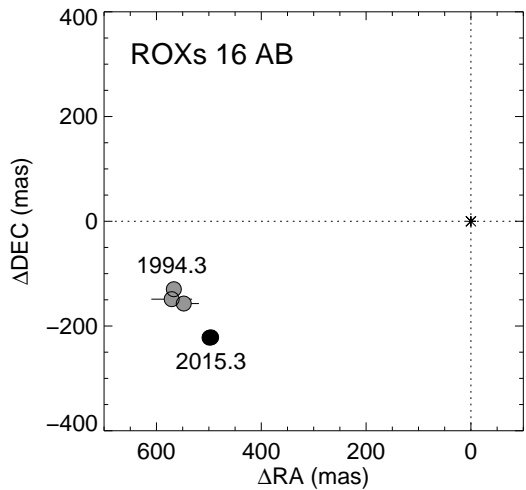}
	\caption{Left: Orbital motion measured for ROXs 16 A,B based on our AO observations (black circles) and previously published measurements in the literature \citep[gray circles;][]{ageorges97,costa00,ratzka05}.}
\label{fig.roxs16_orb}
\end{figure}

ROXs 16 A,B was resolved as a binary with a separation of $\sim$ 590 mas through speckle and shift-and-add techniques \citep{ageorges97,costa00,ratzka05}.  A possible close companion (ROXs 16 Aa,Ab) was detected during the speckle observations by \citet{ratzka05}, although the authors caution that the ``probable third component of ROXs 16 is too close to distinguish between an elongated structure and a point source''.  Fitting the individual Keck AO images as a triple system resulted in possible separations for the close pair ranging between 10$-$20 mas over a wide range of position angles.  Given the large range of positions below the limit of what we can reliably measure using a simultaneous PSF at Keck \citep[e.g.,][]{schaefer12}, we do not consider the detection reliable.  Figure~\ref{fig.roxs16_orb} shows the orbital motion observed in the wide A,B pair.  A statistical analysis of the orbital motion indicates that the orbital period is greater than 42 yr, with the 1\,$\sigma$ confidence interval extending out to the search range of 500 yr.  As shown in Figure~\ref{fig.fratio}, the $K$-band flux ratio of ROXs 16 B relative to A has changed significantly over time.

\subsection{ROXs 42C}

\begin{figure}
	\plotone{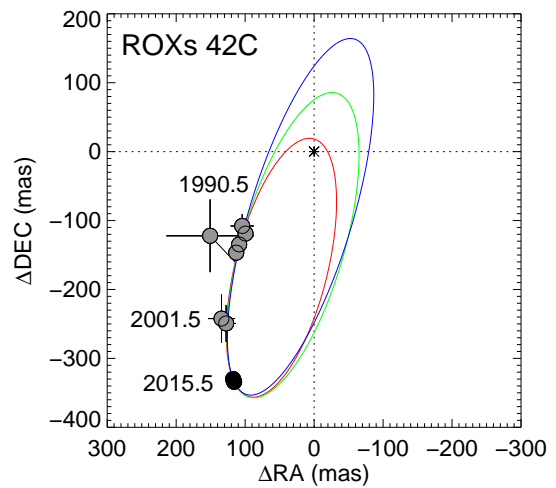}
	\caption{Orbital motion measured for ROXs 42C based on our AO observations (black circles) and published measurements in the literature \citep[gray circles;][]{ghez93,ghez95,aspin97,barsony03,ratzka05}.  Overplotted are examples of orbital solutions with periods ranging from 80 to 120 yr.}
\label{fig.roxs42c_orb}
\end{figure}

ROXs 42C (NTTS 162814-2427) was identified as a double-lined spectroscopic binary with a period of 36 days by \citet{mathieu89}.  It was discovered to have a wide tertiary component with a separation of 150 mas through speckle interferometry \citep{ghez93}.  The orbital motion of the visual pair has been monitored through speckle imaging and shift-and-add techniques \citep{ghez93,ghez95,aspin97,barsony03,ratzka05}.  We resolved the system during two AO observations.  Figure~\ref{fig.roxs42c_orb} shows the orbital motion for ROXs 42C.  A statistical analysis of orbital solutions that fit the data indicates that the orbital period is greater than 56 yr, with the 1\,$\sigma$ confidence interval extending out to the search range of 500 yr.

\subsection{ROXs 43 Aa,Ab,B}

\begin{figure}
	\plotone{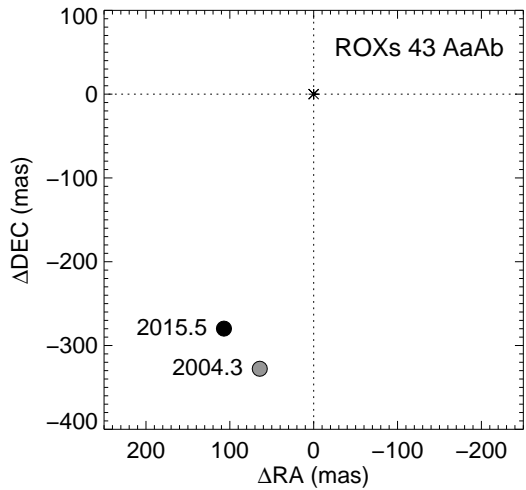}
	\caption{Orbital motion measured for ROXs 43 Aa,Ab based on our AO observation (black circle) and the AO observation by \citet[gray circle;][]{correia06}.}
\label{fig.roxs43a_orb}
\end{figure}

ROXs 43 A and B are a $\sim$ 4\farcs5 visual pair \citep{barton51,bouvier92,reipurth93,simon95,ageorges97,ratzka05,correia06,mccabe06}.  ROXs 43A was detected as a single-lined spectroscopic binary with a period of 89 days \citep{mathieu89}.  The spectroscopic binary was discovered to have an additional companion with a separation of 334 mas and a flux ratio of 0.05 through adaptive optics observations by \citet{correia06}.  This companion was not detected in the speckle observations by \citet{ghez93} or \citet{ratzka05}.  ROXs 43B was discovered to have a close companion with a projected separation of 16 mas through the lunar occultation observations by \citet{simon95}.  Therefore, the system is a quintuple.  During our AO observations at Keck in 2015, we resolved the 300 mas pair ROXs 43 Aa,Ab; the orbital motion is shown in Figure~\ref{fig.roxs43a_orb}.  We attempted to fit ROXs 43 Ba,Bb as a close binary using ROXs 43 Aa as the PSF, but could not derive a reliable separation ($\rho <$ 20 mas).

\subsection{ROXs 47A}

ROXs 47A (DoAr 51) was resolved as a triple system by \citet{barsony03}.  The wide pair has a separation of $\sim$ 790 mas \citep{barsony03,ratzka05,mccabe06}.  The orbital motion of the close pair has been resolved through speckle interferometry, aperture masking, and VLBA radio observations \citep{barsony03,cheetham15,rizzuto16,ortizleon17}.  An orbit for the close pair with a period of 8.2 yr and a semi-major axis of 33 mas was computed by \citet{rizzuto16}.  Based on VLBA radio observations, \citet{ortizleon17} revised the relative orbit and computed the astrometric center of mass motion and a geometric parallax of 6.983 $\pm$ 0.050 mas (143.2 $\pm$ 1.0 pc).  This astrometric solution provided individual masses of 0.791 $\pm$ 0.014 $M_\odot$ and 0.781$\pm$ 0.042 $M_\odot$ for the components in the close pair.  

\begin{figure}
	\plotone{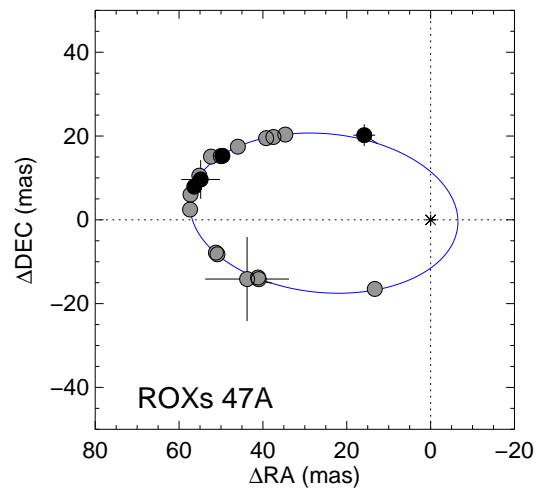}
	\caption{Orbital motion measured for ROXs 47 Aa,Ab (DoAr 51) based on our AO observations (black circles) and previously published measurements \citep[gray circles;][]{barsony03,cheetham15,rizzuto16,ortizleon17}.  Overplotted is the best fitting visual orbit with a period of 8.1 yr.}
\label{fig.roxs47a_orb}
\end{figure}

We resolved the triple system on four nights between 2004$-$2015 using adaptive optics imaging at Keck.  Figure~\ref{fig.roxs47a_orb} shows the orbital motion of the close pair.  Our AO observations are consistent with the previously published measurements.  An orbit fit to all of the available data yields the orbital parameters in Table~\ref{tab.orbpar} which are consistent with those published by \citet{rizzuto16} and \citet{ortizleon17}.  The astrometric motion of the close pair relative to the wide component based on our AO observations is too sparse to model the center of mass motion, however, the motion is consistent with the astrometric orbit published by \citet{ortizleon17}.

As shown in Figure~\ref{fig.fratio}, the $K$-band flux ratio of ROXs 47 Ac relative to Aa and Ab varied by a factor of $\sim$ 2 in 2014$-$2015, while the flux ratio of Ab relative to Aa has changed by only factor of $\sim$ 1.2.  This suggests that Ac is likely the most variable component.

\subsection{EM* SR 12}

\begin{figure}
	\plotone{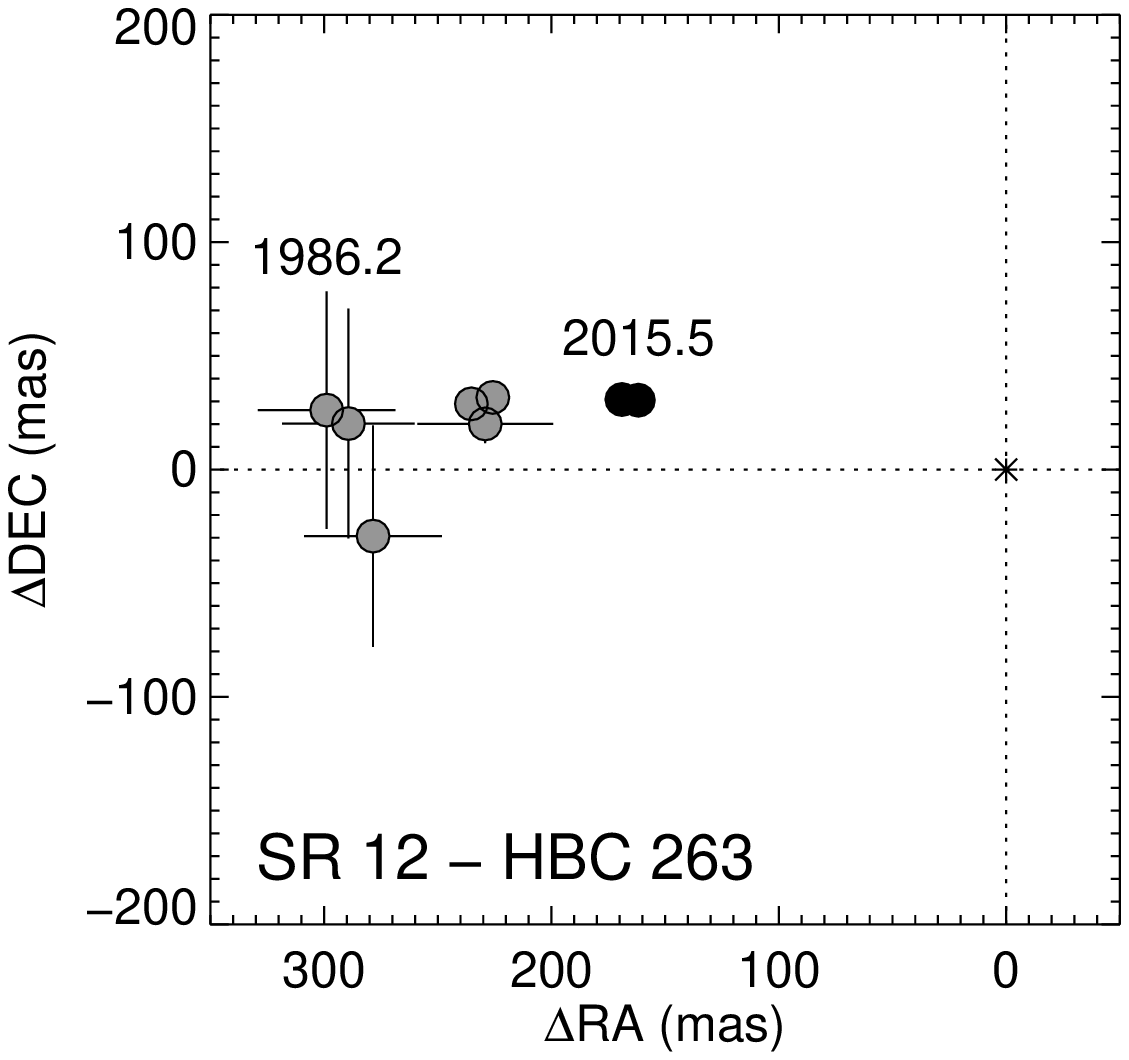}
	\caption{Orbital motion measured for EM* SR 12 A,B based on our AO observations (black circles) and previously published measurements \citep[gray circles;][]{simon87,zinnecker88,simon95,costa00,allen02,mccabe06}. }
\label{fig.sr12_orb}
\end{figure}

SR 12 (HBC 263) was first detected as a binary during a lunar occultation \citep{simon87}.  The orbital motion of the binary has been mapped through lunar occultations, speckle interferometry, and imaging \citep{simon87,zinnecker88,simon95,costa00,allen02,mccabe06}.  The system also has a faint, substellar companion (SR 12 C) at a separation of 8\farcs8 \citep{allen02,kuzuhara11,bowler14}.  SR 12 C was outside of the field of view of the NIRC2 images.  Figure~\ref{fig.sr12_orb} show the orbital motion measured for SR 12 B relative to A.  Fitting an orbit to the data provides an improved $\chi^2$ compared with linear motion, but the coverage is so limited that we cannot place strong constraints on the orbital parameters.  A statistical analysis of orbits that fit the data indicates that the orbital period must be greater than 62 yr with the 1\,$\sigma$ confidence interval extending out to the search range of 500 yr.  The $K$-band flux ratio of SR 12 has increased by a factor of $\sim 1.3$ since the previous measurements in 1996-2001.

\subsection{EM* SR 20}

\begin{figure}
	\plotone{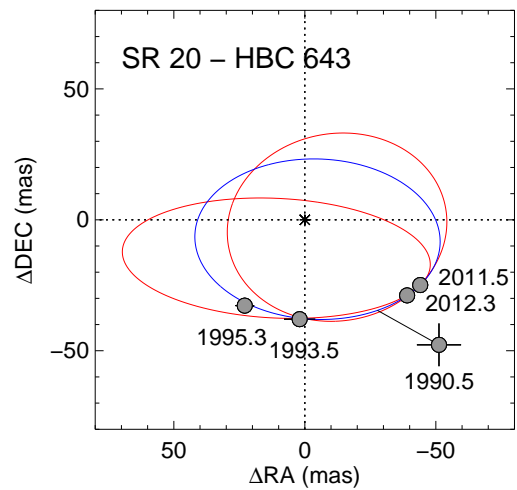}
	\caption{Orbital motion measured for SR 20 (HBC 643) as measured by \citet{ghez93,ghez95} and \citet{cheetham15}.  The blue line shows the best fit orbit with a period of 23.3 yr while the red lines shows orbits within the 1\,$\sigma$ confidence interval in the inclination.}
\label{fig.sr20_orb}
\end{figure}

SR 20 (HBC 643) was first detected as a binary with a separation of 71 mas by \citet{ghez93}.  The orbital motion has been mapped through speckle imaging and aperture masking \citep{ghez93,ghez95,cheetham15}.  The system was unresolved on three occasions reported by \citet{richichi94} and \citet{ghez95}.  We observed the system using AO imaging on UT 2014 July 6; the data suggest that the image is not a single star, but without a simultaneous PSF in the field of view we could not measure a reliable binary separation.  The separation measured from individual frames varied between 20-54 mas over a wide range of position angles.  Likewise, we were not able to measure a reliable separation during aperture masking observations that we obtained on UT 2015 July 12.  As shown in Figure~\ref{fig.sr20_orb}, the published measurements span nearly one full orbital period ($P \sim$ 23.2 yr).  We fit an orbit to these data and obtained the preliminary orbital parameters in Table~\ref{tab.orbpar}.  However, because the observations all lie on the same side of the orbit and there is a large deviation in the position of the discovery observation, most of the orbital parameters have large uncertainties.  The uncertainties were estimated by exploring the 1\,$\sigma$ confidence intervals using the grid search procedure described in \citet{schaefer06}.  At an assumed distance of 130 pc, a statistical analysis of the orbits that fit the data indicate a median total mass of $0.7^{+1.2}_{-0.3}$ $M_\odot$.

\subsection{EM* SR 24N}

\begin{figure}
	\plotone{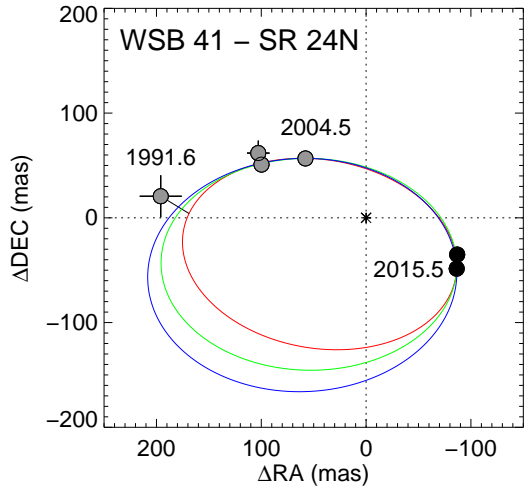}
	\caption{Orbital motion measured for SR 24 Na,Nb based on our AO observations (black circles) and previously published measurements \citep[gray circles;][]{simon95,correia06,mccabe06}.  Overplotted are three examples of orbits that fit the data with orbital periods of 80$-$110 yr.}
\label{fig.sr24n_orb}
\end{figure}

SR 24 S and N are separated by 6$''$ \citep{struve49,ratzka05,mccabe06,correia06}.  Our AO observations indicate that the relative separation has remained constant over the past two decades.  SR 24 N (WSB 41) is a close binary with a separation of $\sim$ 100 mas \citep{simon95,costa00,correia06,mccabe06}.  \citet{cheetham15} did not resolve the close pair during their aperture masking observations.  We resolved all three components in the SR 24 system during two AO imaging observations in 2014$-$2015.  Figure~\ref{fig.sr24n_orb} shows the orbital motion observed for the close binary SR 24 N.  A statistical analysis of orbital solutions that fit the data indicates that the 1\,$\sigma$ confidence interval on the orbital period extends from 78 to 216 yr.  The preliminary ranges for each of the orbital parameters are listed in Table~\ref{tab.orbpar}.  Sufficient curvature has been mapped to provide well-defined 1\,$\sigma$ bounds on the total mass.  Assuming that the system is located in Lynds 1688 at a distance of 137.3 pc, the total mass of SR 24N is 1.24 $\pm$ 0.24 $M_\odot$.  

\begin{figure}
	\plotone{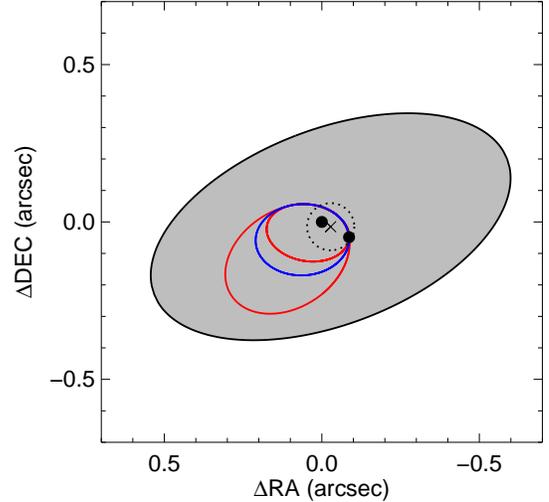}
	\caption{Relative orientation of the SR 24 Na,Nb binary orbit compared with the circumbinary disk.  The filled circles show the relative separation of Na and Nb in 2015.5; the position of Na is fixed at the origin.  The blue line shows the best fit binary orbit with a period of 111 yr while the red lines show the smallest and largest orbits in the 1\,$\sigma$ confidence interval.  The cross marks the relative offset of the continuum emission detected by \citet{fernandezlopez17} in 2015 compared with the infrared positions of Na and Nb relative to SR 24S.  The small dotted circle represents the ALMA resolution limit of 0.15$''$ during the continuum observations.  The shaded ellipse shows the orientation and size of the CO disk in the plane of the sky.}
\label{fig.sr24n_disk}
\end{figure}

The relative brightness of the two components in SR 24N has varied significantly, with the $K$-band flux ratios (Nb/Na) varying between 0.46 to 1.5, as shown in Figure~\ref{fig.fratio}.  There are disks around both SR 24S \citep[e.g.,][]{pinilla17} and SR 24N \citep{greene94,martin98}.  Recent ALMA observations indicate that the disks are misaligned \citep{fernandezlopez17}.  The continuum emission detected around SR 24N is unresolved by the ALMA observations at a resolution of 150 mas.  Based on the 1\,$\sigma$ confidence intervals from the orbital fitting, the minimum and maximum projected separations between Na and Nb range between 42--47 mas and 175--370 mas, suggesting that the continuum emission is likely from a circumstellar disk around one of the northern components.  The semi-major axis of 600 mas for the CO line emission detected by ALMA is larger than the dimensions of the binary orbit, suggesting that it is likely from a circumbinary disk.  This geometry is similar to the interpretation by \citet{andrews05}.  The size and orientation of the circumbinary disk and the binary orbit on the plane of the sky are shown in Figure~\ref{fig.sr24n_disk}.  Comparing the preliminary visual orbital parameters for the close binary with the orientation of the disk around SR 24N indicates a relative of alignment \citep[e.g.,][]{fekel81,kellogg17} of $37^\circ$$^{+19}_{-9}$ or $96^\circ$$^{+17}_{-21}$ between the binary orbit and the disk (the two possibilities are a result of the 180$^\circ$ ambiguity in $\Omega$ for the visual orbit).  Obscuration by the disk around SR 24N could be causing the observed photometric variability; monitoring the flux ratio in the $K$ and $L$ bands could help identify the source of the variability \citep{prato97,beck04}.

\subsection{V853 Oph}

The orbital motion of the wide pair V853 Oph A,B with a separation of $\sim$ 300 mas has been monitored using speckle interferometry, lunar occultation, shift-and-add techniques, the Fine Guidance Sensors onboard the {\it Hubble Space Telescope}, and adaptive optics imaging \citep{ghez93,simon95,aspin97,costa00,mccabe06,schaefer06}.  \citet{simon95} discovered a close companion to the primary (Aa,Ab) during a lunar occultation observation.  We obtained three additional observations of the system using AO imaging at Keck in 2014$-$2015.  On UT 2014 Jul 6, there were significant residuals in the PSF fitting when modeling V853 Oph A as a single source, indicating that the PSF of V853 Oph A is elongated in comparison to B.  For these data, we modeled V853 Oph Aa,Ab as a close binary using V853 Oph B as the PSF and obtained a consistent separation of 23.51 $\pm$ 0.97 mas.  The difference in Akaike's information criterion \citep{burnham02} between the binary and triple fits ranged between 520$-$1000 for the individual images, indicating a signficant improvement in the $\chi^2$.  For our data from the two nights in 2015, we attempted to fit the system as a triple, but the range of separations derived from the individual images varied by a factor of two (14$-$27 mas on 2015 April 5 and 9$-$17 mas on 2015 July 12).  The difference in Akaike's information criterion on these nights was significant (82$-$302), indicating an elongation compared with a point source, however the separation and position angle could not be measured reliably.

\begin{figure}
	\plotone{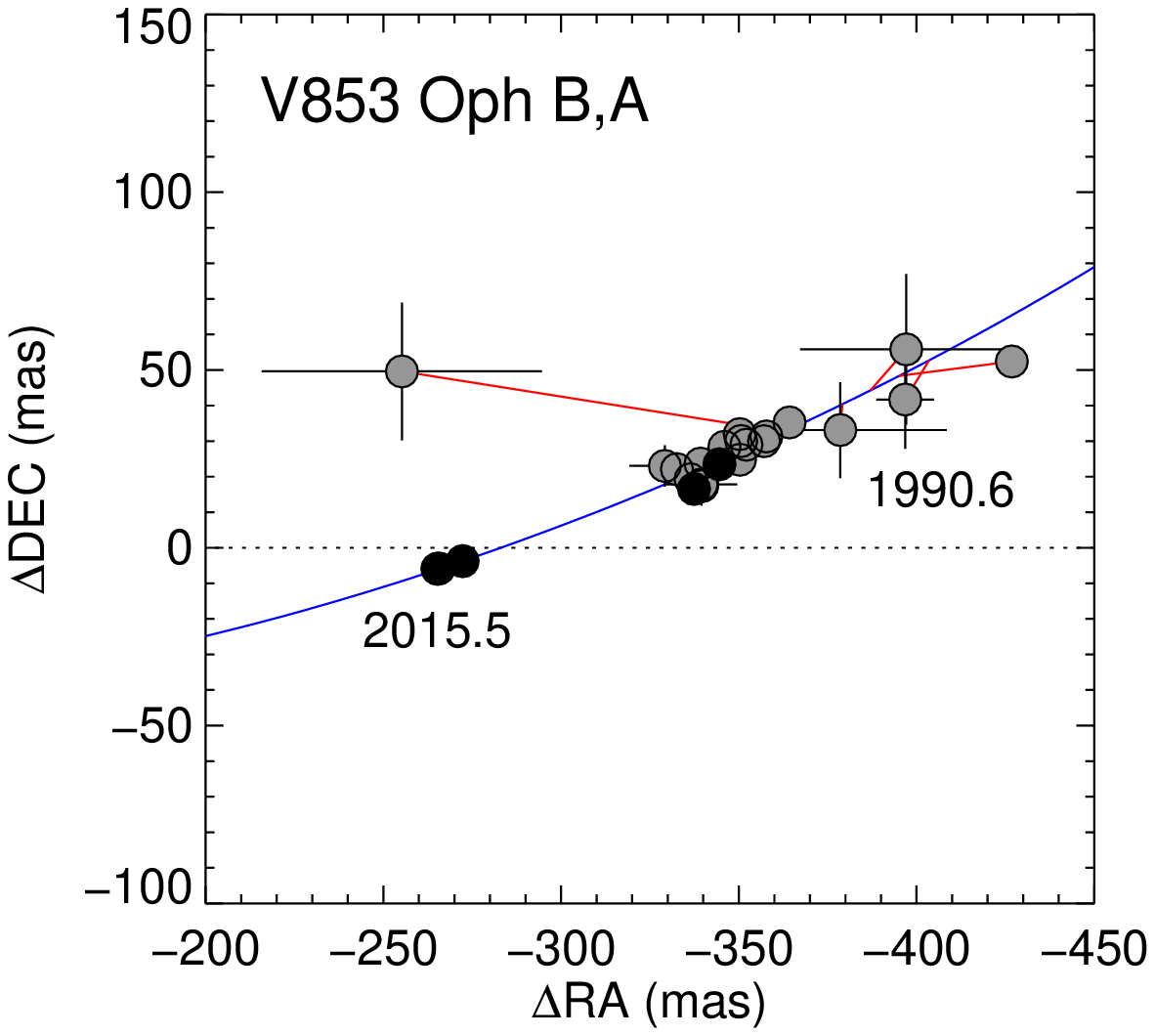}
	\caption{Orbital motion measured for V853 Oph A relative to the fainter wide component B based on our AO observations (black circles) and previously published measurements \citep[gray circles;][]{ghez93,simon95,aspin97,costa00,mccabe06,schaefer06}.  Overplotted in blue is a sample orbit with a period of 400 yr.  The red line segments connect the measurements with the predicted positions along the orbit.  The measured separation of the most discrepant point was near the resolution limit of the telescope in the observed $N$-band \citep{mccabe06}.}
\label{fig.v853oph_orb}
\end{figure}

In Table~\ref{tab.sepPA} we present the wide pair (V853 Oph B,A) separations for all nights, and the possible detection of the close companion in 2014.  We also present two updated positions of the wide pair from \citet{schaefer06} that have been corrected for geometric distortions.  Figure~\ref{fig.v853oph_orb} shows the orbital motion of V853 Oph A relative to the fainter wide component B.  A statistical analysis of possible orbits that fit the data indicates that the period is longer than 80 yr, with the 1\,$\sigma$ confidence interval extending out to the search range of 500 yr.  As shown in Figure~\ref{fig.fratio}, the $K$-band flux ratio of V853Oph A relative to B has varied significantly.  \citet{schaefer06} showed that most of the $V$-band variability is associated with the primary V853 Oph A. Both wide components are classified as classical T Tauri stars with disks \citep{geoffray01}.

\subsection{VSSG 14}

\begin{figure}
	\plotone{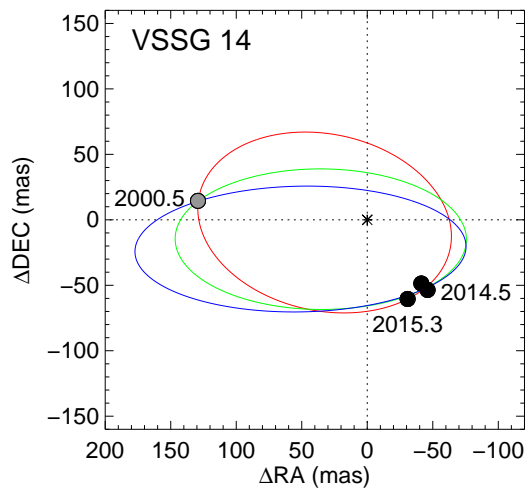}
	\caption{Orbital motion measured for VSSG 14 based on our AO observations (black circles) and the speckle measurement (gray circle) from \citet{ratzka05}.  Overplotted are examples of orbital solutions with periods ranging from 25$-$50 yr.}
\label{fig.vssg14_orb}
\end{figure}

The close companion to VSSG 14 was discovered through a lunar occultation with a projected separation of 101 mas \citep{richichi94}.  It was resolved through speckle imaging eight years later \citep{ratzka05}.  We resolved the close companion during three AO observations in 2014--2015.  Figure~\ref{fig.vssg14_orb} shows the current orbital motion observed for VSSG 14.  A statistical analysis of orbital solutions that fit the existing data indicates that the orbital period is greater than 8.7 yr, with the 1\,$\sigma$ confidence interval extending out to the search range of 500 yrs.  There is not yet enough curvature to place reasonable estimates on the total mass.

\subsection{WSB 18}

\begin{figure}
	\plotone{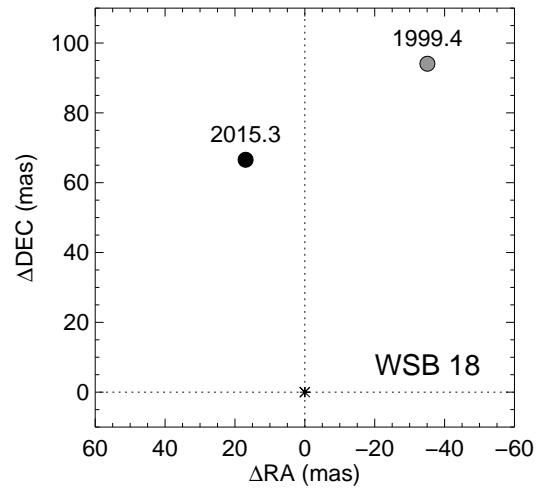}
	\caption{Orbital motion measured for the close pair in WSB 18 based on our AO observation (black circle) and the speckle measurement (gray circle) from \citet{koresko02}.}  

\label{fig.wsb18_orb}
\end{figure}

WSB 18 is a triple system.  The wide pair is separated by 1.0$''$ \citep{reipurth93,koresko02,ratzka05,mccabe06}.  A close companion with a separation of 100 mas was discovered by \citet{koresko02} during speckle observations.  We resolved all three components in the triple system during our AO observation in 2015.  Figure~\ref{fig.wsb18_orb} shows the orbital motion observed for the close pair.  The companion has moved significantly since its detection in 1999.

\subsection{WSB 21}

\begin{figure}
	\plotone{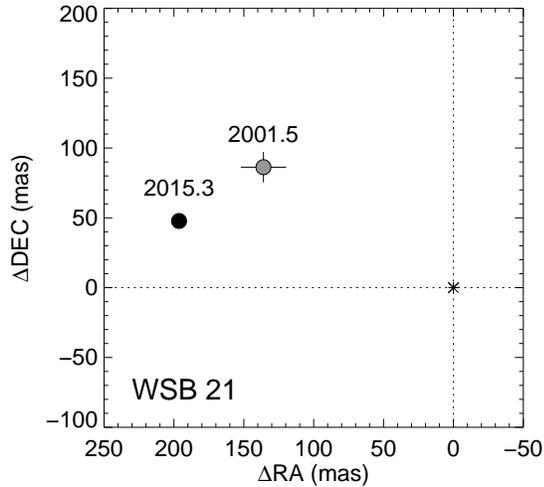}
	\caption{Orbital motion measured for WSB 21 based on our AO observations (black circle) and the speckle measurement (gray circle) from \citet{ratzka05}.}
\label{fig.wsb21_orb}
\end{figure}

\begin{figure}
	\plotone{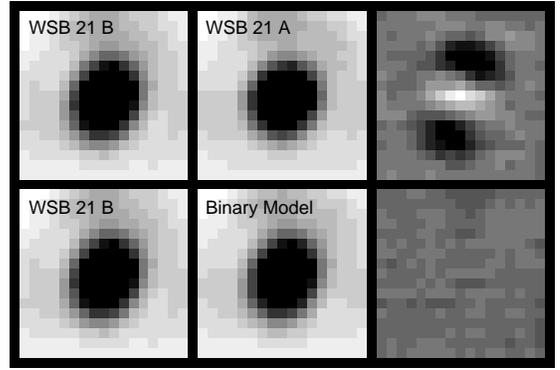}
	\caption{Comparison of binary and triple fits for WSB 21.  The top row shows the fainter component (WSB 21 B), the brighter component (WSB 21 A) scaled to the brightness of WSB 21 B, and the residuals between the images.  The residuals show that WSB 21 B is elongated in comparison to A.  The bottom row shows WSB 21 B, a binary model using WSB 21 A as a simultaneous PSF, and the residuals between the images showing an improved fit.  In both rows, the scale of the residual plots was shifted and expanded by a factor of four to bring out the finer details.} 
\label{fig.compare}
\end{figure}

WSB 21 was discovered to be a binary with a separation of 121 mas through speckle observations by \citet{ratzka05}.  We resolved the companion during our AO observations in 2015.  Figure~\ref{fig.wsb21_orb} shows the orbital motion.  The PSF fitting of the AO data using the primary as the PSF showed strong residuals in the secondary, indicating that the PSF of the secondary is more elongated than the primary.  We attempted to fit the system as a triple and found a consistent solution across all images and filters by fitting the secondary as a close binary with a separation of only 21 mas.  This possible detection of WSB 21 as a triple must be confirmed in future observations.  We list the separation and position angle for both the binary and triple solutions in Table~\ref{tab.sepPA}.  Figure~\ref{fig.compare} shows a comparison of the binary and triple fits.  The separation of the A and B components are close enough on the detector ($\sim$ 200 mas) that we do not expect geometric distortion to be the cause of the difference in the PSF shapes.

\subsection{WSB 37}

\begin{figure}
	\plotone{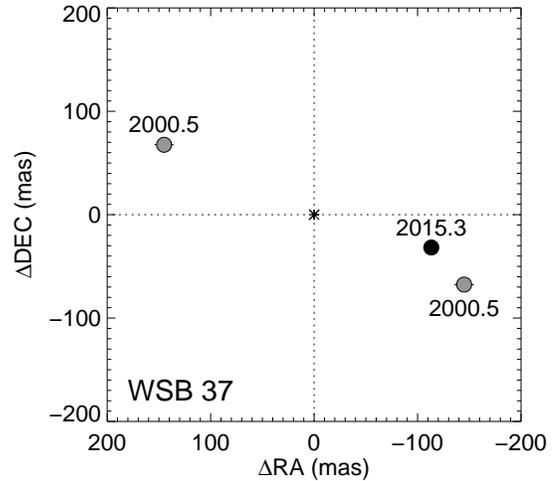}
	\caption{Orbital motion measured for the close pair in WSB 37 based on our AO observations (black circle) and the speckle measurement (gray circle) from \citet{ratzka05}.  There is a 180$^\circ$ ambiguity in the position angle of the speckle observation so we plot both possibilities.}
\label{fig.wsb37_orb}
\end{figure}

WSB 37 was resolved as a binary with a separation of 161 mas through speckle observations by \citet{ratzka05}.  We resolved the system during our AO observations in 2015.  There is a 180$^\circ$ ambiguity in the position angle of the original speckle observation.  The direction of motion could be clarified with additional measurements along the orbit.  Figure~\ref{fig.wsb37_orb} shows two possibilities for how much the binary has moved over fifteen years since its discovery.  The option at a position angle of 245$^\circ$ is more consistent in a dynamical sense with a pair of low-mass stars.  The $K$-band flux ratio of WSB 37 has increased by nearly a factor of 10 compared with the initial measurement by \citet{ratzka05}.

\subsection{WSB 59}

\begin{figure}
	\plotone{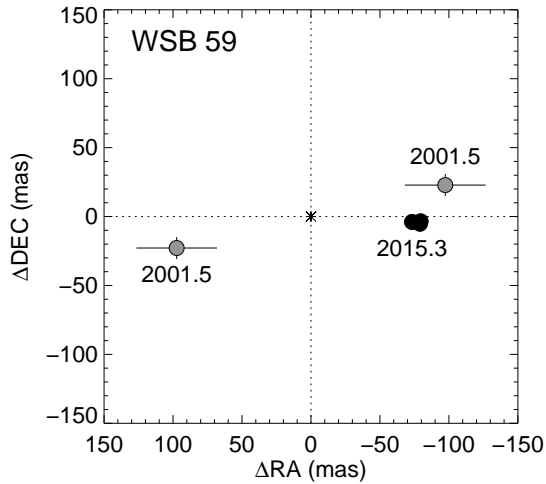}
	\caption{Orbital motion measured for the close pair in WSB 59 based on our AO observations (black circle) and the speckle measurement (gray circle) from \citet{ratzka05}.  The published position angle of the speckle measurement is almost 180$^\circ$ different from the most recent AO observations.  \citet{ratzka05} do not mention explicitly an ambiguity in the position angle but their measurement was near the resolution limit, so we include the flipped position angle in the plot as an alternate possibility.}
\label{fig.wsb59_orb}
\end{figure}

WSB 59 was resolved as a binary with a separation of 100 mas at a position angle of 103$^\circ$ through speckle observations by \citet{ratzka05}.  We resolved the binary during three of our AO observations in 2014$-$2015, most recently with a separation of 73 mas at a position angle of 267$^\circ$.  This suggests a large change in position over the past 15 years.  \citet{ratzka05} do not mention explicitly an ambiguity in the position angle of the speckle observations of WSB 59, although the measured separation is near their resolution limit so they could not distinguish between an elongated structure and a binary star.  If we flip the position angle of that measurement by 180$^\circ$, then the change in position is much smaller.  Figure~\ref{fig.wsb59_orb} shows both possibilities.  Additional measurements along the orbit would clarify the direction of motion.  The $K$-band flux ratio of WSB 59 is twice as large compared with the initial measurement by \citet{ratzka05}.

\section{Conclusions}

We present recent measurements of the orbital positions of 17 multiple systems in the Ophiuchus star forming region and the UCL cluster based on AO imaging at the Keck Observatory.  We detected visual companions of MML 50 and MML 53 for the first time and report the possible detection of a close third companion in WSB 21 and V853 Oph.  We did not resolve EM* SR 20 during our AO observations, but present a preliminary orbit based on previously published measurements.  We also provide a preliminary orbit for EM* SR 24N and a revised visual orbit for the close pair in ROXs 47A based on additional AO measurements reported in this paper.  Most of the observed binaries do not have enough orbital measurements to determine a preliminary orbital solution, however, the positions reported here will help to plan the frequency of future observations and provide data for the eventual determination of the orbital solutions and dynamical masses.

\acknowledgements

We thank the staff at the Keck Observatory for their support during our NIRC2 observing runs.  We thank the referee for providing feedback that improved the paper.  GHS acknowledges support from NASA Keck PI Data Awards administered by the NASA Exoplanet Science Institute and from NSF Grant AST-1411654.  Data presented herein were obtained at the W. M. Keck Observatory from telescope time allocated to the National Aeronautics and Space Administration through the agency's scientific partnership with the California Institute of Technology and the University of California.  The Observatory was made possible by the generous financial support of the W. M. Keck Foundation.  We recognize and acknowledge the significant cultural role that the summit of Mauna Kea plays within the indigenous Hawaiian community and are grateful for the opportunity to conduct these observations from the mountain.  This research has made use of the SIMBAD database and the VizieR catalog access tool, CDS, Strasbourg, France.

%\clearpage

\bibliography{ms}

% Bibliography is treated as a float not as a text, so 
% latex will not insert a page break.  Put invisible text
% on page using empty \mbox command and then call \clearpage.
\mbox

\clearpage

\onecolumngrid
\begin{deluxetable}{lllclcc}
\tabletypesize{\scriptsize}
\tablecaption{Observed Multiple Systems \label{tab.targets}}
\tablewidth{0pt}
\tablehead{
\colhead{Name} & \colhead{RA (2000)} & \colhead{DEC (2000)} & \colhead{Distance (pc)} & \colhead{Region} & \colhead{SpT} & \colhead{References}}
\startdata
MML 50            & 14 50 25.814 & -35 06 48.60 & 140   & UCL    & K0      & S1, D1 \\ 
MML 53            & 14 58 37.697 & -35 40 30.44 & 140   & UCL    & K2      & S1, D1 \\
WSB 18            & 16 24 59.741 & -24 56 00.83 & 130   & Oph    & M3.5    & S2, D2 \\  
WSB 21            & 16 25 15.22  & -25 11 54.1  & 130   & Oph    & M5      & S3, D2 \\
ROX 1-SR 2        & 16 25 19.244 & -24 26 52.68 & 130   & Oph    & G9      & S4, D2 \\ 
ROXs 5            & 16 25 55.825 & -23 55 09.97 & 130   & Oph    & K7      & S5, D2 \\
ROXR1 12          & 16 26 01.61  & -24 29 45.0  & 137.3 & L 1688 & M3      & S6, D3 \\
WSB 37            & 16 26 41.257 & -24 40 17.96 & 137.3 & L 1688 & M5      & S6, D3 \\
ROXs 16           & 16 26 46.43  & -24 12 00.1  & 137.3 & L 1688 & G3.5    & S6, D3 \\  
WSB 41 (SR 24N)   & 16 26 58.44  & -24 45 31.9  & 137.3 & L 1688 & K8      & S7, D3 \\  
HBC 263 (SR 12)   & 16 27 19.512 & -24 41 40.39 & 137.3 & L 1688 & M0      & S6, D3 \\
VSSG 14           & 16 27 49.87  & -24 25 40.2  & 137.3 & L 1688 & A7      & S6, D3 \\
WSB 59            & 16 28 09.209 & -23 52 20.51 & 137.3 & Oph    & M0      & S3, D3 \\
HBC 643 (SR 20)   & 16 28 32.66  & -24 22 44.9  & 130   & Oph    & G7      & S6, D2 \\
V853 Oph          & 16 28 45.28  & -24 28 19.0  & 130   & Oph    & M3.75   & S6, D2 \\ 
ROXs 42C          & 16 31 15.745 & -24 34 02.23 & 147.3 & L 1689 & K6      & S8, D3 \\
ROXs 43A          & 16 31 20.120 & -24 30 05.08 & 147.3 & L 1689 & G0      & S5, D3 \\
ROXs 43B          & 16 31 20.195 & -24 30 00.91 & 147.3 & L 1689 & K5      & S5, D3 \\
ROXs 47A (DoAr51) & 16 32 11.794 & -24 40 21.37 & 143.2 & L 1689 & K3      & S5, D3 \\
\enddata
\tablerefs{Spectral Types: S1 - \citet{mamajek02}; S2 - \citet{erickson11}; 
S3 - Prato et al.\ in prep; S4 - \citet{torres06}; S5 - \citet{bouvier92}; 
S6 - \citet{wilking05}; S7 - \citet{luhman03}; S8 - \citet{cieza10}; Distances: 
D1 - \citet{dezeeuw99}; D2 - \citet{cheetham15}; D3 - \citet{ortizleon17}}
\end{deluxetable} 

\clearpage

\onecolumngrid
\begin{deluxetable}{llllclll}
\tabletypesize{\scriptsize}
\tablecaption{Log of Keck NIRC2 AO Observations \label{tab.aolog}}
\tablewidth{0pt}
\tablehead{
\colhead{UT Date} & \colhead{UT} & \colhead{Target} & \colhead{Filter} & \colhead{AO Rate} & \colhead{$T_{\rm int}$ (s)\tablenotemark{a}} & \colhead{No. Images\tablenotemark{b}} & \colhead{PSF Used}}
\startdata
2004 Jul 14  & 09:15 & ROXs 47A        & $H$, $K'$                 &  50  & 0.18         & 10, 10       & faint wide comp     \\ % 26 mas close pair? - possibly elongated
2014 Jul 05  & 06:36 & MML 53          & $K'$                      & 149  & 4.0          & 30           & ROXs 47B            \\ % AO Rate - fading in and out --- Tertiary companions - deep search - PSF at diff ao rate
             & 07:01 & ROX 1 - SR2     & $J$, $H$, $K'$            & 149  & 1.0--2.0     & 6, 12, 12    & ePSF                \\ % PSF ROXs 47B could not be observed at same AO frame rate - not enough light (clouds?) 
             & 07:24 & ROXs 47B        & $J$, $H$, $K'$            &  60  & 1.0--2.0     & 6, 12, 12    & Single PSF          \\ % Diff AO Rate -  Not enough light to go to higher frame rate
             & 07:43 & ROXs 42C        & $J$, $H$, $K'$            & 149  & 0.2--1.0     & 6, 12, 12    & primary             \\
             & 08:02 & ROXs 5          & $J$, $H$, $K'$, $L'$      & 149  & 0.18--1.0    & 6, 12, 12, 6 & ePSF                \\ % mis-matched PSF shape - use primary instead?  (primary or ROXs 3 or ePSF)
             & 08:23 & ROXs 3          & $J$, $H$, $K'$, $L'$      & 149  & 0.18         & 6, 6, 6, 6   & Single PSF          \\
             & 08:37 & WSB 59          & $K'$                      & 149  & 1.0          & 6            & ePSF                \\ % ROXs 3 - clouds - not enough counts for this frame rate - PSF match isn't great
             & 08:51 & VSSG 14         & $H$, $K'$                 &  60  & 0.5          & 6, 6         & ROXs 3              \\ % AO Rate - clouds - low frame rate (went to Polaris after) - PSF in different conditions
             & 09:46 & HBC 263 - SR 12 & $H$, $K'$                 & 149  & 1.0, 0.5     & 12, 12       & ePSF                \\ % Primary - clouds clearing - ROXs 3 probably not a good PSF!
             &       & HBC 263 - SR 12 & $H$                       & 438  & 0.5          & 6            & ePSF                \\ % Primary - (clouds cleared more?)
2014 Jul 06  & 08:08 & HBC 643 - SR 20 & Hcont, Kcont              & 250  & 0.5--1.0     & 12, 12       & Haro 1-16           \\ % not resolved
             & 08:31 & Haro 1-16       & Hcont, Kcont              & 250  & 1.0          & 6, 6         & Single PSF          \\ % counts up, better AO frame rate (originally 100 Hz)
             & 08:45 & VSSG 14         & Hcont, Kcont              & 149  & 0.5          & 12, 12       & ROXs 10A            \\
             & 08:57 & ROXs 10A        & Hcont, Kcont              & 148  & 0.5          & 6, 6         & Single PSF          \\
             & 09:07 & WSB 59          & Kcont                     &  60  & 1.0--2.0     & 12           & ePSF                \\ % ROXs 10A - AO Rate - Target has pretty low counts
             & 09:20 & ROXs 16         & $K'$                      &  60  & 0.5          & 12           & Tertiary            \\
             & 09:28 & ROXs 47A        & $K'$                      & 149  & 0.5          & 12           & Tertiary            \\
             & 09:38 & WSB 41 - SR 24N & Kcont                     &  60  & 0.5          & 12           & SR 24S              \\
             & 09:49 & V853 Oph        & $H$, Hcont, Kcont         & 250  & 0.2--1.0     & 6, 6, 6      & Tertiary            \\
2015 Apr 05  & 11:01 & MML 53          & Jcont, Hcont, Kcont, $L'$ & 438  & 0.18--1.0    & 6, 12, 12, 6 & MML 50 A            \\
             & 11:22 & MML 50          & Jcont, Hcont, Kcont, $L'$ & 438  & 0.18--1.0    & 6, 6, 6, 6   & Primary             \\ % Used as PSF - binary!
             & 11:41 & WSB 18          & $J$, $H$, $K'$, $L'$      & 440  & 0.18--1.0    & 6, 12, 12, 6 & Wide component      \\ % Triple
             & 12:09 & ROXs 16         & Jcont, Hcont, Kcont, $L'$ & 440  & 0.18--1.0    & 6, 12, 12, 6 & Faint tertiary      \\ % Do we believe 14 mas close pair? Can't see by eye.
             & 12:34 & WSB 21          & $J$, $H$, $K'$, $L'$      & 440  & 0.18--0.5    & 6, 12, 12, 6 & Primary             \\ % possible triple - possible elongation
             & 12:52 & ROXs 20A        & $J$, $H$, $K'$, $L'$      & 440  & 0.18--0.5    & 6, 6, 6, 6   & Single PSF          \\
             & 13:10 & WSB 37          & $J$, $H$, $K'$, $L'$      & 440  & 0.18--0.5    & 6, 12, 12, 6 & ePSF                \\ % ROXs 20A -- ePSF
             & 13:40 & WSB 59          & $J$, $H$, $K'$, $L'$      & 750  & 0.18--0.5    & 6, 12, 12, 6 & ePSF                \\ % ROXs 20A -- ePSF
             & 13:59 & ROXs 20A        & $J$, $H$, $K'$, $L'$      & 750  & 0.18--0.5    & 6, 12, 6, 6  & Single PSF          \\
             & 14:18 & ROXR1 12        & $J$, $H$, $K'$, $L'$      & 750  & 0.18--0.5    & 6, 12, 12, 6 & ePSF                \\ % ROXs 20A -- ePSF
             & 14:40 & VSSG 14         & Jcont, Hcont, Kcont, $L'$ & 750  & 0.18--2.0    & 6, 12, 12, 6 & ROXs 20B            \\
             & 14:59 & ROXs 20B        & Jcont, Hcont, Kcont, $L'$ & 750  & 0.18--2.0    & 6, 6, 6, 6   & Single PSF          \\
             & 15:12 & V853 Oph        & Jcont, Hcont, Kcont, $L'$ & 750  & 0.18--0.5    & 6, 12, 12, 6 & Tertiary            \\
             & 15:29 & ROXs 47A        & $J$, $H$, Kcont, $L'$     & 750  & 0.18--0.5    & 6, 12, 12, 6 & Tertiary            \\
2015 Jul 12  & 05:46 & MML 53          & Hcont, Kcont              & 438  & 0.5--1.0     & 12, 12       & MML 50 A            \\ 
             & 05:59 & MML 50          & Hcont, Kcont              & 438  & 0.5--0.7     & 6, 6         & MML 50 A            \\ % PSF - wide binary
             & 06:13 & ROXs 43 A       & Jcont, Hcont, Kcont, $L'$ & 438  & 0.18--0.6  & 10, 10, 10, 20 & ROXs 43 Aa          \\ % Aa,Ab----B  (BaBb using Aa as PSF - not resolved)
             & 07:33 & ROXs 5          & Jcont, Hcont, Kcont, $L'$ & 250  & 0.18--1.5    & 6, 6, 6, 6   & ePSF                \\ % ROXs 3 -- ePSF
             & 07:50 & ROXs 3          & Jcont, Hcont, Kcont, $L'$ & 250  & 0.18--1.0    & 6, 6, 6, 6   & Single PSF          \\
             & 08:05 & HBC 263 - SR 12 & Jcont, Hcont, Kcont, $L'$ & 250  & 0.18--1.0    & 6, 6, 6, 6   & ePSF                \\ % ROXs 3 -- ePSF
             & 08:21 & ROXs 42C        & Jcont, Hcont, Kcont, $L'$ & 438  & 0.18--0.5    & 6, 6, 6, 6   & Primary             \\
             & 08:37 & V853 Oph        & Hcont, Kcont              & 250  & 0.5--0.8     & 12, 12       & Wide faint comp     \\ % Close pair 13 mas - pretty hard to say
             & 08:47 & ROX 1 - SR 2    & Jcont, Hcont, Kcont, $L'$ & 438  & 0.18--0.8    & 6, 6, 6, 6   & ePSF    \\
             & 09:00 & ROXs 47B        & Jcont, Hcont, Kcont, $L'$ & 438  & 0.18--1.0    & 3, 3, 6, 3   & Single PSF          \\ 
             & 09:16 & WSB 41 - SR 24N & Jcont, Hcont, Kcont, $L'$ & 149  & 0.18--0.5  & 10, 10, 10, 10 & SR 24S              \\ % Triple
             & 09:30 & ROXs 47A        & Hcont, Kcont              & 250  & 0.5          & 5, 5         & Faint tertiary      \\ % Triple
\enddata
\tablenotetext{a}{Integration time per co-add.  Each image is composed of 10 co-added exposures.  If different exposure times were used for different sets of images, then the range of values is listed.}
\tablenotetext{b}{The number of images taken in each filter.}
\end{deluxetable}

%\clearpage

\onecolumngrid
\begin{deluxetable}{lllll} 
%\tabletypesize{\small}
\tabletypesize{\scriptsize}
\tablewidth{0pt}
\tablecaption{Keck NIRC2 Adaptive Optics Measurements of the Orbital Positions in Multiple Systems\tablenotemark{$a$} \label{tab.sepPA}} 
\tablecolumns{5}
\tablehead{
\colhead{Julian Year} & \colhead{$\rho$(mas)} & \colhead{P.A.($\degr$)} & \colhead{Filter} & \colhead{Flux Ratio}}
\startdata 
\multicolumn{5}{c}{MML 50} \\
 \noalign{\vskip .8ex}%
 \hline
 \noalign{\vskip .8ex}%
2015.2580  &   375.58  $\pm$   0.66  &   20.906  $\pm$  0.101  & Jcont & 0.0919 $\pm$ 0.0039 \\
           &                         &                         & Hcont & 0.1029 $\pm$ 0.0010 \\
           &                         &                         & Kcont & 0.1266 $\pm$ 0.0038 \\
           &                         &                         & Lcont & 0.1427 $\pm$ 0.0025 \\
2015.5257  &   375.59  $\pm$   0.28  &   21.043  $\pm$  0.044  & Hcont & 0.1038 $\pm$ 0.0016 \\
           &                         &                         & Kcont & 0.1267 $\pm$ 0.0013 \\
\cutinhead{MML 53}
2014.5073  &    57.07  $\pm$   2.46  &    19.16  $\pm$   2.47  & $K'$  & 0.530 $\pm$ 0.116  \\ % - fading in and out
2015.2579  &    53.67  $\pm$   1.93  &    28.34  $\pm$   2.06  & Jcont & 0.429 $\pm$ 0.045  \\
           &                         &                         & Hcont & 0.416 $\pm$ 0.030  \\
           &                         &                         & Kcont & 0.389 $\pm$ 0.020  \\
           &                         &                         & $L'$  & 0.370 $\pm$ 0.035  \\
2015.5256  &    50.58  $\pm$   1.30  &    36.49  $\pm$   1.48  & Hcont & 0.359 $\pm$ 0.016  \\
           &                         &                         & Kcont & 0.403 $\pm$ 0.017  \\
\cutinhead{ROX 1 - EM* SR 2}
2014.5073  &   153.69  $\pm$   2.08  &  125.507  $\pm$  0.774  & $J$   & 1.064 $\pm$ 0.131  \\ % epsf
           &                         &                         & $H$   & 1.005 $\pm$ 0.016  \\ % epsf
           &                         &                         & $K'$  & 1.010 $\pm$ 0.044  \\ % epsf
2015.5260  &   149.50  $\pm$   0.13  &  123.474  $\pm$  0.054  & Jcont & 0.9720 $\pm$ 0.0031 \\ % epsf - PA = 123.47392
           &                         &                         & Hcont & 0.9639 $\pm$ 0.0021 \\ % epsf - PA = 123.47392
           &                         &                         & Kcont & 0.9624 $\pm$ 0.0020 \\ % epsf - PA = 123.47392
           &                         &                         & $L'$  & 0.9568 $\pm$ 0.0033 \\ % epsf - PA = 123.47392
\cutinhead{ROXR1 12}
2015.2583  &   107.04  $\pm$   0.52  &     9.99  $\pm$   0.28  & $J$   & 0.8218 $\pm$ 0.0279 \\ % epsf
           &                         &                         & $H$   & 0.7837 $\pm$ 0.0075 \\ % epsf
           &                         &                         & $K'$  & 0.8031 $\pm$ 0.0050 \\ % epsf
           &                         &                         & $L'$  & 0.8466 $\pm$ 0.0179 \\ % epsf
\cutinhead{ROXs 5}
2014.5074  &   184.07  $\pm$   0.35  &   351.40  $\pm$   0.11  & $J$   & 0.3921 $\pm$ 0.0070 \\ % epsf
           &                         &                         & $H$   & 0.4309 $\pm$ 0.0044 \\ % epsf
           &                         &                         & $K'$  & 0.4387 $\pm$ 0.0072 \\ % epsf
           &                         &                         & $L'$  & 0.4490 $\pm$ 0.0220 \\ % epsf
2015.5258  &   180.43  $\pm$   0.34  &   352.96  $\pm$   0.11  & Jcont & 0.3845 $\pm$ 0.0087 \\ % epsf
           &                         &                         & Hcont & 0.4200 $\pm$ 0.0049 \\ % epsf
           &                         &                         & Kcont & 0.4435 $\pm$ 0.0037 \\ % epsf
           &                         &                         & $L'$  & 0.4616 $\pm$ 0.0033 \\ % epsf
\cutinhead{ROXs 16 A,B} % - binary fit
2014.5103  &   545.99  $\pm$   0.35  &  114.017  $\pm$  0.037  & $K'$  & 0.1915 $\pm$ 0.0011 \\
2015.2581  &   543.05  $\pm$   0.71  &  114.093  $\pm$  0.076  & Jcont & 0.2670 $\pm$ 0.0218 \\
           &                         &                         & Hcont & 0.2152 $\pm$ 0.0042 \\
           &                         &                         & Kcont & 0.1572 $\pm$ 0.0023 \\
           &                         &                         & $L'$  & 0.0884 $\pm$ 0.0009 \\
\cutinhead{ROXs 42C}
2014.5074  &   350.29  $\pm$   0.34  &  160.414  $\pm$  0.056  & $J$   & 0.2515 $\pm$ 0.0028 \\
           &                         &                         & $H$   & 0.2640 $\pm$ 0.0038 \\
           &                         &                         & $K'$  & 0.2554 $\pm$ 0.0028 \\
2015.5259  &   353.58  $\pm$   0.31  &  160.855  $\pm$  0.053  & Jcont & 0.2460 $\pm$ 0.0031 \\
           &                         &                         & Hcont & 0.2534 $\pm$ 0.0018 \\
           &                         &                         & Kcont & 0.2603 $\pm$ 0.0012 \\
           &                         &                         & $L'$  & 0.2455 $\pm$ 0.0020 \\
\cutinhead{ROXs 43 Aa,Ab}
2015.5257  &   299.61  $\pm$   0.75  &   159.07  $\pm$   0.14  & Jcont & 0.0335 $\pm$ 0.0016 \\
           &                         &                         & Hcont & 0.0394 $\pm$ 0.0012 \\
           &                         &                         & Kcont & 0.0402 $\pm$ 0.0009 \\
           &                         &                         & $L'$  & 0.0470 $\pm$ 0.0014 \\
\cutinhead{ROXs 43 Aa,B}
2015.5257  &  4479.6   $\pm$   1.2   &   11.880  $\pm$  0.025  & Jcont & 0.595 $\pm$ 0.029 \\
           &                         &                         & Hcont & 0.625 $\pm$ 0.024 \\
           &                         &                         & Kcont & 0.538 $\pm$ 0.011 \\
           &                         &                         & $L'$  & 0.364 $\pm$ 0.018 \\
\cutinhead{ROXs 47 Aa,Ab}
2004.5336  &    25.67  $\pm$   2.60  &    38.08  $\pm$   5.80  & $H$   & 1.166 $\pm$ 0.309  \\ % ff1/ff2:  0.91678021  0.25157478
           &                         &                         & $K'$  & 1.159 $\pm$ 0.408  \\ % ff1/ff2:  0.98801514  0.43532945
2014.5103  &    51.95  $\pm$   0.73  &    72.90  $\pm$   0.81  & $K'$  & 0.896 $\pm$ 0.012  \\
2015.2584  &    55.68  $\pm$   4.61  &    80.04  $\pm$   4.75  & $J$   & 0.858 $\pm$ 0.052  \\
           &                         &                         & $H$   & 0.834 $\pm$ 0.080  \\
           &                         &                         & Kcont & 0.807 $\pm$ 0.097  \\
           &                         &                         & $L'$  & 0.912 $\pm$ 0.022  \\
2015.5261  &    56.92  $\pm$   0.60  &    82.01  $\pm$   0.61  & Hcont & 0.967 $\pm$ 0.037  \\ % ff1/ff2:  1.0358383  0.041115761 
           &                         &                         & Kcont & 0.979 $\pm$ 0.008  \\ % ff1/ff2:  1.0216947  0.0077401844
\cutinhead{ROXs 47 Ac,Aa}
2004.5336  &   796.57  $\pm$   2.26  &  260.205  $\pm$  0.163  & $H$   & 3.404 $\pm$ 0.463  \\ % 1/ff1:  0.29863166  0.040110377
           &                         &                         & $K'$  & 2.475 $\pm$ 0.478  \\ % 1/ff1:  0.41701160  0.076541613
2014.5103  &   815.57  $\pm$   0.55  &  264.554  $\pm$  0.040  & $K'$  & 3.340 $\pm$ 0.036  \\ % 1/ff1:  0.29940219  0.0032147679
2015.2584  &   818.04  $\pm$   4.35  &  265.040  $\pm$  0.305  & $J$   & 2.665 $\pm$ 0.205  \\ % 1/ff1:  0.37703882  0.029112259
           &                         &                         & $H$   & 2.403 $\pm$ 0.142  \\ % 1/ff1:  0.41762374  0.026473269
           &                         &                         & Kcont & 1.685 $\pm$ 0.093  \\ % 1/ff1:  0.59520369  0.033805944 
           &                         &                         & $L'$  & 1.106 $\pm$ 0.015  \\ % 1/ff1:  0.90449542  0.011985127
2015.5261  &   820.21  $\pm$   1.00  &  265.172  $\pm$  0.073  & Hcont & 2.484 $\pm$ 0.107  \\ % 1/ff1:  0.40322762  0.016620709 
           &                         &                         & Kcont & 1.646 $\pm$ 0.026  \\ % 1/ff1:  0.60750610  0.0094959775
\cutinhead{ROXs 47 Ac,Ab}
2004.5336  &   777.72  $\pm$   2.75  &  261.474  $\pm$  0.203  & $H$   & 3.843 $\pm$ 0.532  \\ % 1/ff2:  0.26490273  0.038007092
           &                         &                         & $K'$  & 2.697 $\pm$ 0.502  \\ % 1/ff2:  0.38461659  0.085408618
2014.5103  &   764.76  $\pm$   0.49  &  265.341  $\pm$  0.038  & $K'$  & 2.993 $\pm$ 0.025  \\ % 1/ff2:  0.33417499  0.0027783066
2015.2584  &   762.58  $\pm$   3.82  &  265.405  $\pm$  0.287  & $J$   & 2.279 $\pm$ 0.098  \\ % 1/ff2:  0.43945770  0.019081095
           &                         &                         & $H$   & 1.993 $\pm$ 0.070  \\ % 1/ff2:  0.50223603  0.017158799
           &                         &                         & Kcont & 1.354 $\pm$ 0.121  \\ % 1/ff2:  0.74443869  0.071002104 
           &                         &                         & $L'$  & 1.008 $\pm$ 0.012  \\ % 1/ff2:  0.99195516  0.012016551
2015.5261  &   763.38  $\pm$   1.10  &  265.408  $\pm$  0.085  & Hcont & 2.398 $\pm$ 0.043  \\ % 1/ff2:  0.41718153  0.0075400900
           &                         &                         & Kcont & 1.611 $\pm$ 0.018  \\ % 1/ff2:  0.62064199  0.0069980967
\cutinhead{EM* SR 12 (HBC 263)}
2014.5076  &   171.80  $\pm$   0.22  &   79.669  $\pm$  0.074  & H     & 0.9373 $\pm$ 0.0078 \\ % ePSF
           &                         &                         & $K'$  & 0.9357 $\pm$ 0.0049 \\ % ePSF
2015.5259  &   164.62  $\pm$   0.17  &   79.326  $\pm$  0.062  & Jcont & 0.9506 $\pm$ 0.0060 \\ % ePSF
           &                         &                         & Hcont & 0.9699 $\pm$ 0.0055 \\ % ePSF
           &                         &                         & Kcont & 0.9502 $\pm$ 0.0054 \\ % ePSF
           &                         &                         & $L'$  & 0.9414 $\pm$ 0.0049 \\ % ePSF
\cutinhead{EM* SR 24 Na,Nb (WSB 41)}
2014.5103  &    93.73  $\pm$   1.58  &   248.03  $\pm$   0.97  & Kcont & 0.461 $\pm$ 0.023 \\
2015.5260  &    99.11  $\pm$   0.29  &   240.68  $\pm$   0.17  & Jcont & 0.973 $\pm$ 0.030 \\
           &                         &                         & Hcont & 0.992 $\pm$ 0.020 \\
           &                         &                         & Kcont & 1.059 $\pm$ 0.017 \\
           &                         &                         & $L'$  & 1.091 $\pm$ 0.008 \\
\cutinhead{EM* SR 24 S,Na}
2014.5103  &  5071.40  $\pm$  10.28  &   349.48  $\pm$   0.12  & Kcont & 0.585 $\pm$ 0.041 \\
2015.5260  &  5070.38  $\pm$   8.94  &   349.48  $\pm$   0.10  & Jcont & 0.504 $\pm$ 0.024 \\
           &                         &                         & Hcont & 0.470 $\pm$ 0.021 \\
           &                         &                         & Kcont & 0.464 $\pm$ 0.014 \\
           &                         &                         & $L'$  & 0.519 $\pm$ 0.036 \\
\cutinhead{EM* SR 24 S,Nb}
2014.5103  &  5053.63  $\pm$   9.85  &   348.44  $\pm$   0.11  & Kcont & 0.270 $\pm$ 0.026 \\
2015.5260  &  5039.31  $\pm$   9.02  &   348.42  $\pm$   0.10  & Jcont & 0.490 $\pm$ 0.020 \\
           &                         &                         & Hcont & 0.467 $\pm$ 0.027 \\
           &                         &                         & Kcont & 0.492 $\pm$ 0.019 \\
           &                         &                         & $L'$  & 0.567 $\pm$ 0.043 \\
\cutinhead{V853 Oph B,A}
2003.2855  &   345.37  $\pm$   1.85  &  273.901  $\pm$  0.307  & $H$   & 2.945 $\pm$ 0.064 \\ %  1/ff1:  0.33976201  0.0074480910
           &                         &                         & $K'$  & 3.484 $\pm$ 0.105 \\ %  1/ff1:  0.28731537  0.0088505576
2004.5335  &   337.81  $\pm$   0.39  &  272.811  $\pm$  0.066  & $H$   & 2.823 $\pm$ 0.039 \\ %  1/ff1:  0.35435045  0.0048995992
           &                         &                         & $K'$  & 3.362 $\pm$ 0.056 \\ %  1/ff1:  0.29755030  0.0050074027
2014.5104  &   272.34  $\pm$   0.52  &  269.209  $\pm$  0.109  & $H$   & 3.305 $\pm$ 0.043 \\ %  1/ff1:  0.30261786  0.0039645134
           &                         &                         & Hcont & 3.219 $\pm$ 0.061 \\ %  1/ff1:  0.31075656  0.0060480691
           &                         &                         & Kcont & 3.634 $\pm$ 0.034 \\ %  1/ff1:  0.27518525  0.0025771752
2015.2584  &   265.67  $\pm$   0.62  &  268.744  $\pm$  0.135  & Jcont & 2.498 $\pm$ 0.471 \\ %  1/ff1:  0.41357968  0.084724455
           &                         &                         & Hcont & 3.457 $\pm$ 0.327 \\ %  1/ff1:  0.29238879  0.036202505
           &                         &                         & Kcont & 4.312 $\pm$ 0.057 \\ %  1/ff1:  0.23194195  0.0030817393
           &                         &                         & $L'$  & 5.527 $\pm$ 0.138 \\ %  1/ff1:  0.18103434  0.0045979355
2015.5260  &   265.26  $\pm$   0.12  &  268.723  $\pm$  0.032  & Kcont & 5.109 $\pm$ 0.018 \\ %  1/ff1:  0.19573882  0.0006998274
           &                         &                         & Hcont & 4.136 $\pm$ 0.023 \\ %  1/ff1:  0.24179826  0.0013199004
\cutinhead{V853 Oph B,Aa - possible triple fit}
2014.5104  &   276.95  $\pm$   1.00  &   271.09  $\pm$   0.21  & $H$   & 1.943 $\pm$ 0.033 \\ %  1/ff1: 0.51480402  0.008676252
           &                         &                         & Hcont & 1.992 $\pm$ 0.044 \\ %  1/ff1: 0.50230281  0.011215369
           &                         &                         & Kcont & 2.297 $\pm$ 0.080 \\ %  1/ff1: 0.43575538  0.015416653
\cutinhead{V853 Oph B,Ab - possible triple fit}
2014.5104  &   265.69  $\pm$   0.60  &   266.73  $\pm$   0.13  & $H$   & 1.591 $\pm$ 0.058 \\ %  1/ff2: 0.62933075  0.023036143
           &                         &                         & Hcont & 1.505 $\pm$ 0.149 \\ %  1/ff2: 0.67018505  0.070346394
           &                         &                         & Kcont & 1.679 $\pm$ 0.093 \\ %  1/ff2: 0.59696180  0.032891194
\cutinhead{V853 Oph Aa,Ab - possible triple fit}
2014.5104  &    23.51  $\pm$   0.97  &   150.34  $\pm$   2.36  & $H$   & 0.819 $\pm$ 0.036 \\
           &                         &                         & Hcont & 0.757 $\pm$ 0.091 \\
           &                         &                         & Kcont & 0.733 $\pm$ 0.064 \\
\cutinhead{VSSG 14}
2014.5075\tablenotemark{$b$}  &    63.77  $\pm$   2.22  &   220.34  $\pm$   1.99  & $K'$  & 0.415 $\pm$ 0.035 \\ % K ONLY - H poor quality
2014.5102  &    70.62  $\pm$   1.63  &   220.68  $\pm$   1.32  & Hcont & 0.286 $\pm$ 0.022 \\
           &                         &                         & Kcont & 0.365 $\pm$ 0.010 \\
2015.2583  &    67.85  $\pm$   2.93  &   206.99  $\pm$   2.47  & Jcont & 0.163 $\pm$ 0.037 \\
           &                         &                         & Hcont & 0.240 $\pm$ 0.032 \\
           &                         &                         & Kcont & 0.338 $\pm$ 0.044 \\
           &                         &                         & $L'$  & 0.377 $\pm$ 0.013 \\
\cutinhead{WSB 18 Aa,Ab}
2015.2580  &    68.69  $\pm$   0.72  &    14.28  $\pm$   0.60  & $J$   & 0.554 $\pm$ 0.024 \\ % PSF not as stable at J
           &                         &                         & $H$   & 0.506 $\pm$ 0.021 \\
           &                         &                         & $K'$  & 0.511 $\pm$ 0.008 \\
           &                         &                         & $L'$  & 0.539 $\pm$ 0.047 \\ % Close pair not visually separated at L-band
\cutinhead{WSB 18 B,Aa}
2015.2580  &  1097.41  $\pm$   0.92  &   259.324 $\pm$  0.049  & $J$   & 1.184 $\pm$ 0.031 \\ % PSF not as stable at J
           &                         &                         & $H$   & 1.178 $\pm$ 0.043 \\
           &                         &                         & $K'$  & 0.968 $\pm$ 0.016 \\
           &                         &                         & $L'$  & 0.631 $\pm$ 0.020 \\ % Close pair not visually separated at L-band
\cutinhead{WSB 18 B,Ab}
2015.2580  &  1070.23  $\pm$   0.97  &   262.659 $\pm$  0.053  & $J$   & 0.655 $\pm$ 0.021 \\ % PSF not as stable at J
           &                         &                         & $H$   & 0.596 $\pm$ 0.013 \\
           &                         &                         & $K'$  & 0.495 $\pm$ 0.011 \\
           &                         &                         & $L'$  & 0.339 $\pm$ 0.022 \\ % Close pair not visually separated at L-band
\cutinhead{WSB 21 A,B - binary fit}
2015.2581  &   202.04  $\pm$   0.36  &    76.33  $\pm$   0.10  & $J$   & 0.8918 $\pm$ 0.0092 \\ % primary as PSF
           &                         &                         & $H$   & 0.8037 $\pm$ 0.0057 \\ % primary as PSF
           &                         &                         & $K'$  & 0.7721 $\pm$ 0.0026 \\ % primary as PSF
           &                         &                         & $L'$  & 0.6772 $\pm$ 0.0131 \\ % primary as PSF
\cutinhead{WSB 21 Ba,Bb - possible triple fit}
2015.2581  &    21.47  $\pm$   0.80  &   164.09  $\pm$   2.15  & $J$   & 0.74 $\pm$ 0.12 \\ % ff1/ff2
           &                         &                         & $H$   & 0.73 $\pm$ 0.10 \\ % ff1/ff2
           &                         &                         & $K'$  & 0.72 $\pm$ 0.10 \\ % ff1/ff2
           &                         &                         & $L'$  & 0.50 $\pm$ 0.15 \\ % ff2/ff1
\cutinhead{WSB 21 A,Ba - possible triple fit}
2015.2581  &   201.99  $\pm$   0.67  &    73.84  $\pm$   0.19  & $J$   & 0.534 $\pm$ 0.038 \\ % ff2
           &                         &                         & $H$   & 0.486 $\pm$ 0.029 \\ % ff2
           &                         &                         & $K'$  & 0.474 $\pm$ 0.027 \\ % ff2
           &                         &                         & $L'$  & 0.474 $\pm$ 0.048 \\ % ff1
\cutinhead{WSB 21 A,Bb - possible triple fit}
2015.2581  &   203.03  $\pm$   1.00  &    79.91  $\pm$   0.28  & $J$   & 0.391 $\pm$ 0.035 \\ % ff1
           &                         &                         & $H$   & 0.352 $\pm$ 0.030 \\ % ff1
           &                         &                         & $K'$  & 0.341 $\pm$ 0.025 \\ % ff1
           &                         &                         & $L'$  & 0.232 $\pm$ 0.048 \\ % ff2
\cutinhead{WSB 37}
2015.2582  &   117.65  $\pm$   0.16  &  254.354  $\pm$  0.079  & $J$   & 1.0223 $\pm$ 0.0124 \\ % epsf
           &                         &                         & $H$   & 0.9917 $\pm$ 0.0030 \\ % epsf
           &                         &                         & $K'$  & 0.9773 $\pm$ 0.0034 \\ % epsf
           &                         &                         & $L'$  & 0.9755 $\pm$ 0.0086 \\ % epsf
\cutinhead{WSB 59}
2014.5075  &    79.17  $\pm$   5.98  &   266.21  $\pm$   4.33  & $K'$  & 0.566 $\pm$ 0.062 \\ % epsf
2014.5103  &    79.59  $\pm$   3.01  &   267.56  $\pm$   2.17  & Kcont & 0.596 $\pm$ 0.041 \\ % epsf
2015.2582  &    73.36  $\pm$   0.51  &   266.94  $\pm$   0.40  & $J$   & 0.476 $\pm$ 0.013 \\ % epsf
           &                         &                         & $H$   & 0.505 $\pm$ 0.009 \\ % epsf
           &                         &                         & $K'$  & 0.564 $\pm$ 0.007 \\ % epsf
           &                         &                         & $L'$  & 0.648 $\pm$ 0.018 \\ % - ROXs 20A (could not create ePSF)
\enddata 
\tablenotetext{a}{When listed, the ``X,Y'' designation indicates that component Y was measured relative to X.}
\tablenotetext{b}{$H$-band images of VSSG 14 from UT 2014 Jul 5 (2014.5075) were of poor quality and not included in the final results.  The discrepancy between the two positions measured in 2014.5 is likely because of a larger mismatch of PSF shapes on UT 2014 Jul 5.  The components of VSSG 14 were too blended to create an effective PSF.}
\tablecomments{ROX 47 Ac,Aa and Ac,Ab give the position angle of the brighter component relative to the fainter tertiary.  This angle is flipped 180$^\circ$ from the position angle given in the literature for the faint companion relative to the brighter primary.}
\end{deluxetable}

\onecolumngrid
\begin{deluxetable}{lccc} 
%\tabletypesize{\small}
\tabletypesize{\scriptsize}
\tablewidth{0pt}
\tablecaption{Orbital Parameters \label{tab.orbpar}} 
\tablecolumns{3}
\tablehead{
\colhead{Parameters} & \colhead{ROXs 47A} & \colhead{EM* SR 20} & \colhead{EM* SR 24 N}}
\startdata 
$P$ (yr)           &    8.051 $\pm$ 0.039   &   23.2$^{+1.1}_{-0.9}$  & 111$^{+105}_{-33}$ \\
$T$ (JY)           & 2012.035 $\pm$ 0.035   &  2002.6$^{+7.7}_{-5.4}$ & 2009.34$^{+0.15}_{-0.13}$ \\
$e$                &   0.7961 $\pm$ 0.0062  &   0.26$^{+0.73}_{-0.11}$ & 0.64$^{+0.13}_{-0.10}$ \\
$a$ (mas)          &    32.76 $\pm$ 0.48    &    48$^{+420}_{-6}$     & 181$^{+83}_{-30}$ \\
$i$ ($^\circ$)      &     20.7 $\pm$ 4.6     &   130$^{+21}_{-37}$     & 132.1$^{+7.4}_{-5.0}$ \\
$\Omega$ ($^\circ$) &     51.4 $\pm$ 8.5     &    90$^{+43}_{-13}$     & 72.0$^{+4.3}_{-3.4}$ \\
$\omega$ ($^\circ$) &    216.9 $\pm$ 9.0     &    67$^{+41}_{-46}$     & 110.9$^{+5.0}_{-5.8}$ \\
$M_{\rm tot}$ ($M_\odot$) &    1.593 $\pm$ 0.079   & 0.7$^{+1.2}_{-0.3}$ $(\frac{d}{\rm 130 pc})^3$    & 1.24$^{+0.24}_{-0.23}$ $(\frac{d}{\rm 137.3 pc})^3$ \\
\enddata
\tablecomments{The total mass of ROXs 47A includes the uncertainty in the orbital parallax (143.2 $\pm$ 1.0 pc) determined by \citet{ortizleon17}. The total mass of SR 20 is the median from a statistical analysis of orbits that fit the data; the uncertainties contain 34\% of the solutions on either side of the median.  The total mass of SR 24 N was determined from the 1\,$\sigma$ ($\Delta\chi^2$ = 1) confidence interval.  The uncertainties in the masses for SR 20 and SR 24 are derived from the visual orbit only and do not include uncertainties in the distance.  The distances for these systems are not yet precisely known and will be refined by astrometric missions like GAIA.}
\end{deluxetable}

\end{document}